\documentclass[a4paper,nofootinbib,superscriptaddress,
secnumarabic,showpacs,titlepage,eqsecnum]{revtex4}

\usepackage{amsmath, amsthm, amsfonts, amssymb, latexsym}
\usepackage{graphicx}
\usepackage{color}
\usepackage{epsfig}

\usepackage[a4paper]{geometry}

\usepackage[caption=false]{subfig}

\usepackage{url}

\usepackage[unicode]{hyperref}
\hypersetup{
    unicode=true,          
    pdftitle={Numerical relativity for D dimensional axially symmetric
      space-times},
    pdfauthor={Miguel Zilh\~ao, Helvi Witek, Ulrich Sperhake, Vitor Cardoso,
      Leonardo Gualtieri, Carlos Herdeiro, Andrea Nerozzi},
    pdfsubject={Numerical Relativity},
    pdfcreator={LaTeX},          
    pdfproducer={LaTeX},         
    pdfkeywords={General Relativity, Numerical Relativity, Black Holes},
    colorlinks=false,            
  }

\newcommand{\half}{\frac{1}{2}}
\def\bequ{\begin{equation}}
\def\eequ{\end{equation}}

\begin{document}

\title{\Large Numerical relativity for $D$ dimensional axially symmetric
       space-times: formalism and code tests}

\author{Miguel Zilh\~ao}\email{mzilhao@fc.up.pt}
\affiliation{
  Centro de F\'\i sica do Porto --- CFP  \\
  Departamento de F\'\i sica e Astronomia \\
  Faculdade de Ci\^encias da Universidade do Porto --- FCUP \\
  Rua do Campo Alegre, 4169-007 Porto, Portugal
}

\author{Helvi Witek}\email{helvi.witek@ist.utl.pt}
\affiliation{
  Centro Multidisciplinar de Astrof\'\i sica --- CENTRA \\
  Departamento de F\'\i sica, Instituto Superior T\'ecnico --- IST \\
  Av. Rovisco Pais 1, 1049-001 Lisboa, Portugal  
}

\author{Ulrich Sperhake}\email{sperhake@tapir.caltech.edu}
\affiliation{
  California Institute of Technology \\  
  Pasadena, CA 91125, USA
}

\author{Vitor Cardoso}\email{vitor.cardoso@ist.utl.pt}
\affiliation{
  Centro Multidisciplinar de Astrof\'\i sica --- CENTRA \\
  Departamento de F\'\i sica, Instituto Superior T\'ecnico --- IST \\
  Av. Rovisco Pais 1, 1049-001 Lisboa, Portugal 
}
\affiliation{
  Department of Physics and Astronomy, The University of Mississippi \\
  University, MS 38677-1848, USA
}

\author{Leonardo Gualtieri}\email{leonardo.gualtieri@roma1.infn.it}
\affiliation{
  Dipartimento di Fisica, Universit\`a di Roma
  ``Sapienza'' \& Sezione \\
  INFN Roma1, P.A. Moro 5, 00185, Roma, Italy
}

\author{Carlos Herdeiro}\email{crherdei@fc.up.pt}
\affiliation{
  Centro de F\'\i sica do Porto --- CFP \\ 
  Departamento de F\'\i sica e Astronomia \\ 
  Faculdade de Ci\^encias da Universidade do Porto --- FCUP \\
  Rua do Campo Alegre, 4169-007 Porto, Portugal
}

\author{Andrea Nerozzi}\email{andrea.nerozzi@ist.utl.pt}
\affiliation{
  Centro Multidisciplinar de Astrof\'\i sica --- CENTRA \\
  Departamento de F\'\i sica, Instituto Superior T\'ecnico --- IST \\
  Av. Rovisco Pais 1, 1049-001 Lisboa, Portugal 
}

\date{January 2010} 

\begin{abstract}
  The numerical evolution of Einstein's field equations in a generic background
  has the potential to answer a variety of important questions in physics: from
  applications to the gauge-gravity duality, to modelling black hole production
  in TeV gravity scenarios, analysis of the stability of exact solutions and
  tests of Cosmic Censorship. In order to investigate these questions, we extend
  numerical relativity to more general space-times than those investigated
  hitherto, by developing a framework to study the numerical evolution of
  $D$~dimensional vacuum space-times with an $SO(D-2)$ isometry group for $D\ge
  5$, or $SO(D-3)$ for $D\ge 6$.

  Performing a dimensional reduction on a $(D-4)$-sphere, the $D$~dimensional
  vacuum Einstein equations are rewritten as a 3+1 dimensional system with
  source terms, and presented in the Baumgarte, Shapiro, Shibata and Nakamura
  (BSSN) formulation. This allows the use of existing 3+1 dimensional numerical
  codes with small adaptations. Brill-Lindquist initial data are constructed in
  $D$~dimensions and a procedure to match them to our 3+1 dimensional evolution
  equations is given. We have implemented our framework by adapting the
  \textsc{Lean} code and perform a variety of simulations of non-spinning black
  hole space-times. Specifically, we present a modified {\em moving puncture}
  gauge which facilitates long term stable simulations in $D=5$. We further
  demonstrate the internal consistency of the code by studying convergence and
  comparing numerical versus analytic results in the case of geodesic slicing
  for $D=5,6$.

\end{abstract}

\pacs{~04.25.D-,~04.25.dg,~04.50.-h,~04.50.Gh}

\maketitle

\tableofcontents

\section{Introduction}
Numerical relativity is an essential tool to study many processes involving
strong gravitational fields. In four space-time dimensions, processes of this
sort, such as black hole (BH) binary evolutions, are of utmost importance for
understanding the main sources of gravitational waves, which are expected to be
detected by the next generation of ground based [Laser Interferometer
Gravitational-Wave Observatory (LIGO), VIRGO] and space based [Laser
Interferometer Space Antenna (LISA)] interferometers. Long-term stable numerical
evolutions of BH binaries have finally been achieved after four decades of
efforts \cite{Pretorius:2005gq,Campanelli:2005dd,Baker:2005vv}.  The numerical
modelling of generic spinning BH binaries in vacuum Einstein gravity is an
active field of research, with important consequences for gravitational wave
detection in the near future.

Numerical relativity in a higher dimensional space-time, instead, is an
essentially unexplored field, with tremendous potential to provide answers to
some of the most fundamental questions in physics.  Recent developments in
experimental and theoretical physics make this a pressing issue. We refer, in
particular, to the prominent role of BHs in the gauge-gravity duality, in
TeV-scale gravity or even on their own as solutions of the field
equations. These are some of the most active areas of current research in
gravitational and high energy physics.

\subsection{Motivation}

\renewcommand{\theenumi}{\textit{\roman{enumi}}}

\begin{enumerate}

\item \label{item:1} \emph{AdS/CFT and holography.}  In 1997--98, a powerful new
  technique known as the AdS/CFT correspondence or, more generally, the
  gauge-string duality, was introduced and rapidly developed
  \cite{Maldacena:1997re}. This holographic correspondence provides an effective
  description of a non-perturbative, strongly coupled regime of certain gauge
  theories in terms of higher-dimensional classical gravity. In particular,
  equilibrium and non-equilibrium properties of strongly coupled thermal gauge
  theories are related to the physics of higher-dimensional BHs, black branes
  and their fluctuations. These studies revealed intriguing connections between
  the dynamics of BH horizons and hydrodynamics \cite{Son:2007vk}, and offer new
  perspectives on notoriously difficult problems, such as the BH information
  loss paradox, the nature of BH singularities or quantum gravity.

  Numerical relativity in anti-de Sitter backgrounds is bound to contribute
  enormously to our understanding of the gauge-gravity duality and is likely to
  have important applications in the interpretation of observations
  \cite{Mateos:2007ay,Hartnoll:2009sz,Amsel:2007cw,Gubser:2008pc}.  For
  instance, in the context of the gauge-gravity duality, high energy collisions
  of BHs have a dual description in terms of \textit{a)}~high energy collisions
  with balls of de-confined plasma surrounded by a confining phase and
  \textit{b)}~the rapid localised heating of a de-confined plasma. These are the
  type of events that may have direct observational consequences for the
  experiments at Brookhaven's Relativistic Heavy Ion Collider (RHIC)
  \cite{Amsel:2007cw,Gubser:2008pc}.  Numerical relativity in anti-de Sitter is
  notoriously difficult, and so far only very special situations have been
  handled \cite{Pretorius:2000yu,Witek:2010qc}. The phenomenologically most
  interesting case is a five dimensional space-time, $AdS_5$, and therefore the
  higher dimensional extension of numerical relativity is necessary.

\item \label{item:2} \emph{TeV-scale gravity scenarios.}  An outstanding problem
  in high energy physics is the extremely large ratio between the four
  dimensional Planck scale, $10^{19}$ GeV, and the electroweak scale, $10^2$
  GeV. It has been proposed that this \textit{hierarchy problem} can be resolved
  if one adopts the idea that the Standard Model is confined to a brane in a
  higher dimensional space, such that the extra dimensions are much larger than
  the four dimensional Planck scale (they may be large up to a sub-millimetre
  scale) \cite{Antoniadis:1990ew,ArkaniHamed:1998rs,Antoniadis:1998ig}. In a
  different version of the model, the extra dimensions are infinite, but the
  metric has an exponential factor introducing a finite length scale
  \cite{Randall:1999ee,Randall:1999vf}.

  In such models, the fundamental Planck scale
  could be as low as 1 TeV. Thus, high energy colliders, such as the Large
  Hadron Collider (LHC), may directly probe strongly coupled gravitational
  physics \cite{Argyres:1998qn,Banks:1999gd,Giddings:2001bu,Dimopoulos:2001hw,
    Ahn:2002mj,Chamblin:2004zg}.  In fact, such tests may even be routinely
  available in the collisions of ultra-high energy cosmic rays with the Earth's
  atmosphere \cite{Feng:2001ib,Ahn:2003qn,Cardoso:2004zi}, or in astrophysical
  BH environments \cite{Banados:2009pr,Berti:2009bk,Jacobson:2009zg} (for
  reviews see \cite{Cavaglia:2002si,Kanti:2004nr,Kanti:2008eq}).  From Thorne's
  hoop conjecture it follows that, in this scenario, particle collisions could
  produce BHs \cite{Giddings:2001bu,Dimopoulos:2001hw}. Moreover, the production
  of BHs at trans-Planckian collision energies (compared to the fundamental
  Planck scale) should be well described by using classical general relativity
  extended to $D$~dimensions
  \cite{Banks:1999gd,Giddings:2001bu,Dimopoulos:2001hw,Feng:2001ib,Ahn:2003qn,
    Ahn:2002mj,Chamblin:2004zg,Cardoso:2004zi,Cavaglia:2002si,Kanti:2004nr,
    Kanti:2008eq,Solodukhin:2002ui,Hsu:2002bd}.  The challenge is then to use
  the classical framework to determine the cross section for production and, for
  each initial setup, the fractions of the collision energy and angular momentum
  that are lost in the higher dimensional space by emission of gravitational
  waves.  This information will be of paramount importance to improve the
  modelling of microscopic BH production in event generators such as
  \textsc{Truenoir}, \textsc{Charybdis2}, \textsc{Catfish} or \textsc{Blackmax}
  \cite{Dimopoulos:2001hw,Frost:2009cf,Cavaglia:2006uk,Dai:2007ki,Dai:2009by}.
  The event generators will then provide a description of the corresponding
  evaporation phase, which might be observed during LHC collisions.

  The first models for BH production in parton-parton collisions used a simple
  black disk approach to estimate the cross section for production
  \cite{Giddings:2001bu,Dimopoulos:2001hw}. Improved bounds have been obtained
  using either trapped surface methods to estimate the cross section for BH
  production
  \cite{Eardley:2002re,Kohlprath:2002yh,Yoshino:2002br,Yoshino:2002tx} or
  approximation schemes \cite{D'Eath:1992hb,D'Eath:1992hd,D'Eath:1992qu,
    Cardoso:2002ay,Berti:2003si,Cardoso:2005jq} to evaluate the gravitational
  energy loss.  Only recently exact results for highly relativistic collisions
  where obtained in four dimensions, using numerical relativity techniques
  \cite{Sperhake:2008ga,Shibata:2008rq,Sperhake:2009jz}.  No such exact results
  are yet available in the higher dimensional case. To obtain them is one of our
  main goals and the present paper introduces a formalism to achieve that.

\item \label{item:3} \emph{Higher dimensional black holes.}  Asymptotically flat
  higher dimensional black objects have a much richer structure than their four
  dimensional counterparts. For instance, spherical topology is not the only
  allowed topology for objects with a horizon. One can also have, \emph{e.g.},
  black rings, with a donut-like topology. Remarkably, these two different
  horizon topologies coexist for certain regions in phase-space
  \cite{Emparan:2008eg}.  The stability of general higher-dimensional BHs is now
  starting to be explored.  Generically it has been conjectured that for $D\ge
  6$ ultra-spinning Myers-Perry BHs will be unstable \cite{Emparan:2003sy}. This
  instability has been confirmed by an analysis of linearised axi-symmetric
  perturbations in $D=7,8,9$ \cite{Dias:2009iu}. Clearly, the study of the
  non-linear development of these instabilities requires numerical methods, such
  as the ones presented herein. A study of this type was very recently presented
  for a non axi-symmetric perturbation in $D=5$ \cite{Shibata:2009ad}, where it
  was found that a single spinning five dimensional Myers-Perry BH is unstable,
  for sufficiently large rotation parameter (thereby confirming previous
  conjectures \cite{Cardoso:2006sj,Cardoso:2009bv,Cardoso:2009nz}).

  Not much is known about general equilibrium states in anti-de Sitter
  backgrounds. The gauge-gravity duality and the hydrodynamic limit have been
  used to predict the existence of larger classes of BHs in anti-de Sitter
  backgrounds, including non axi-symmetric solutions
  \cite{Cardoso:2009bv,Cardoso:2009nz}. However, these have not yet been found.

\end{enumerate}

Finally, there are issues of principle, as for example testing Cosmic Censorship
in BH collisions \cite{Sperhake:2008ga,Sperhake:2009jz} which require
state-of-the-art numerical simulations.

\subsection{Space-times with symmetries}
From what has been said, the extension of four dimensional numerical Relativity
is mandatory.  Some pioneering works have been concerned with the non-linear
development of the Gregory-Laflamme instability \cite{Gregory:1993vy} of cosmic
strings \cite{Choptuik:2003qd} and gravitational collapse, with spherical
symmetry \cite{Sorkin:2009bc}, axial symmetry \cite{Sorkin:2009wh} or even
static situations \cite{Headrick:2009pv}. Another numerical code, based on the
cartoon method \cite{Alcubierre:1999ab}, was developed and tested for five
space-time dimensions in Ref.~\cite{Yoshino:2009xp}. See also
Ref.~\cite{Nakao:2009dc} for a discussion of slicings of $D$ dimensional black
holes. The (phenomenologically) most interesting large extra dimensions models
are, however, in higher than five space-time dimensions (see for instance
\cite{Kanti:2004nr}). Moreover, the ultra-spinning instabilities of Myers-Perry
BHs should occur in $D\ge 6$. Thus, our approach here is to develop a framework
and a numerical code that can, in principle, be applied to different space-time
dimensions with little adaptations. This may be achieved by taking the
$D$~dimensional vacuum space-time to have an isometry group fit to include a
large class of interesting problems. If this isometry group is sufficiently
large, it allows a dimensional reduction of the problem to 3+1 dimensions,
wherein it appears as (four dimensional) general relativity coupled to some
\textit{quasi-matter} terms.\footnote{Hereafter, we dub the source terms of the
  lower dimensional Einstein equations as \textit{quasi-matter}, since its
  energy-momentum tensor is not that of canonical matter.} Thus, the different
space-time dimension manifests itself only in the different quasi-matter content
of the four dimensional theory.  We emphasise, in this context, that full blown
$4+1$, $5+1$, \emph{etc.\ }numerical simulations without symmetry are currently
not possible due to the computational costs, so that our approach pushes
numerical relativity in higher dimensions to the outmost practical limits of the
present time.  Moreover, an obvious advantage of this approach is that we can
use existing codes with small adaptations: the four dimensional equations need
to be coupled to the appropriate quasi-matter terms and some issues related to
the chosen coordinates must be addressed, as we shall see. Finally, the lessons
learnt in treating our effective gravity plus quasi-matter system might be of
use in dealing with other four dimensional numerical relativity problems with
sources.

\subsection{Axial symmetry $SO(D-2)$ and $SO(D-3)$}
\label{sec:axial-symmetry}
We consider two classes of models, which are generalisations of axial symmetry
to higher dimensional space-times: a $D\ge5$ dimensional vacuum space-time with
an $SO(D-2)$ isometry group, and a $D\ge6$ dimensional vacuum space-time with an
$SO(D-3)$ isometry group. The former class allows studies of head-on collisions
of non-spinning BHs. In order to end up with a $3+1$ dimensional model we use,
however, only part of this symmetry: we perform a dimensional reduction by
isometry on a $(D-4)$-sphere which has an $SO(D-3) \subset SO(D-2)$ isometry
group.  The latter class allows to model BH collisions with impact parameter and
with spinning BHs, as long as all the dynamics take place on a single
plane.\footnote{This follows from the fact that the angular momenta of the black
  holes are parallel to the orbital angular momentum.} In this case we perform a
dimensional reduction by isometry on the entire $SO(D-3)$ isometry group.  This
class includes the most interesting physical configurations relevant to
accelerator---and cosmic ray---physics (in the context of TeV-scale gravity),
and to the theoretical properties of higher-dimensional black objects (such as
stability and phase diagrams).

We formulate the evolution equations in the Baumgarte, Shapiro, Shibata and
Nakamura (BSSN) formulation \cite{Shibata:1995we,Baumgarte:1998te}, together
with the moving puncture approach \cite{Campanelli:2005dd,Baker:2005vv}.  This
is known to provide a stable evolution scheme for vacuum solutions in four
dimensions, and therefore it is the natural framework for our Einstein plus
quasi-matter system.  The quasi-matter terms however, exhibit a problem for
numerical evolution, well known from other numerical studies using coordinates
adapted to axial symmetry, which is sourced by the existence of a coordinate
singularity at the axis. In our formulation, this problem appears when a certain
3+1 dimensional Cartesian coordinate vanishes, $y=0$. We present a detailed
treatment of this problem, introducing first regular variables, then analysing
one by one all potentially pathological terms in our evolution equations and
finally presenting a method to heal all of them. The resulting equations have no
further (obvious) problems for numerical evolution and could, in principle, be
implemented in any working 3+1 dimensional numerical relativity code.

Here we present numerical results using the \textsc{Lean} code
\cite{Sperhake:2006cy}, developed by one of us.  We stress that the formalism
developed here is valid in general $D$.  However, long term stable evolutions
typically require some experiments with free parameters in the gauge conditions
and also possibly with constraint damping. For $D=5$ we show that, if
appropriate gauge conditions are chosen, the numerical evolution for
Brill-Lindquist initial data describing a single BH is stable and the
constraints are preserved in the evolution, within numerical error. As another
test, we evolve the same initial data in a geodesic slicing gauge. This gauge is
inappropriate for a long term evolution; but it allows us to compare the
numerical evolution with the analytic solution for a single Tangherlini BH in
$D=5$. We find excellent agreement between the two. We also present some
preliminary results for $D=6$.

This paper is organised as follows. In Section~\ref{sec:3+1-dimens}, we discuss
the $D$ dimensional ansatz, perform the dimensional reduction by isometry,
perform the Arnowitt-Deser-Misner (ADM) split and present the BSSN formulation
of our equations. In Section~\ref{sec:initial-data}, the construction of
Brill-Lindquist initial data in $D$ dimensions is discussed and a procedure to
match it to our $3+1$ formulation is given. In
Section~\ref{sec:numerical-treatment} we present the numerical treatment and
results. We draw our conclusions and discuss implications of our results for
future work in Section~\ref{sec:final-remarks}. A considerable part of the
technical details for the numerical treatment is organised into three
appendices. In Appendix~\ref{axis} we motivate and discuss the introduction of
regular variables at $y=0$ and present all relevant equations in terms of these
variables. In Appendix~\ref{troubleterms} we explain how to tackle all the
problematic terms at $y=0$ in these equations. Finally, in
Appendix~\ref{geoslice}, we discuss the construction of the geodesic slicing
which is used to compare analytical with numerical results.

\section{The effective 3+1 dimensional system}
\label{sec:3+1-dimens}
The starting point of the formalism used here is a dimensional reduction from
$D$~dimensional general relativity in vacuum to a four dimensional model. The
isometry group of $D$~dimensional Minkowski space-time is $ISO(1,D-1)$;
solutions of general relativity (or of other metric theories of gravity)
generically break this symmetry into a subgroup. For instance, the isometry
group of a Schwarzschild (or, for $D>4$, Tangherlini \cite{Tangherlini:1963bw})
BH is $SO(D-1)\times \mathbb{R}$, whereas for a head-on collision of two
non-rotating BHs it is $SO(D-2)$: indeed, neither the time direction nor the
direction of the collision correspond to symmetries, but a rotation of the
remaining $D-2$ spatial directions leaves the space-time invariant. The total
space-time can then be considered as the semi-direct product of a three
dimensional space-time ${\cal N}$ with the sphere $S^{D-3}=SO(D-2)/SO(D-3)$. A
coordinate system for ${\cal N}$ can be given, for example in the case of a
head-on collision of two BHs, by the time $t$, the coordinate $z$ along the
collision axis, and the distance from that axis.

One can take advantage of this symmetry to reduce the space-time
dimensionality. This can be accomplished by writing Einstein's equations in
$D$~dimensions in a coordinate system which makes the symmetry manifest,
allowing for a lower dimensional interpretation of the $D$~dimensional
Einstein's equations (in the spirit of Kaluza-Klein reduction). We remark,
however, that we are not performing a compactification; rather, we perform a
dimensional reduction by isometry, as first proposed by Geroch
\cite{Geroch:1970nt}. The extra dimensions manifest themselves in the lower
dimensionality as a source of Einstein's equations, defined on the lower
dimensional manifold.

In principle, one could use the symmetry in a more na\"ive way, assuming that
the solution does not depend on the coordinates parameterizing the sphere and
simply evolving the relevant components of the $D$~dimensional Einstein's
equations. The perspective provided by dimensional reduction, however, has two
advantages: \textit{(i)} all quantities have a geometrical interpretation, and
this allows for a deeper understanding of the problem and a better control of
the equations; \textit{(ii)} it is possible to use, with minor modifications,
the numerical codes which have already been written to implement Einstein's
equations in a four dimensional space-time.  Therefore, we do not use the entire
$SO(D-2)$ symmetry of the process, but only a $SO(D-3)$ subgroup. This reduces
the space-time on a $(D-4)$-sphere and yields a four dimensional manifold.

In the original proposal of Geroch \cite{Geroch:1970nt} the symmetry space was
$SO(2)$.  This approach has been applied to numerical relativity, see for
instance \cite{Sjodin:2000zd,Sperhake:2000fe,Choptuik:2003as}; a five
dimensional extension, with the same symmetry space, has been derived in
\cite{Chiang:1985rk}.  A generalisation to coset manifolds (like the sphere
$S^n$) was given by Cho in \cite{Cho:1986wk,Cho:1987jf}, but in these papers the
complete form of Einstein's equations was not presented. Here we provide the
explicit form of Einstein's equations for symmetry spaces $S^n$ together with
their numerical implementation.

\subsection{$4+(D-4)$ split}
We now describe in detail the reduction from $D$~to $4$ dimensions.  In order to
highlight the particular classes of BH binaries we are able to study with this
framework, it is convenient to begin this discussion with the isometry group of
the $S^{D-3}$ sphere, i.~e.~with the $3+(D-3)$ split.

A general $D$~dimensional space-time metric may be written in the form
\bequ
  d\hat{s}^2=\hat{g}_{MN}dx^Mdx^N=g_{\bar{\mu}\bar{\nu}}(x^M)dx^{\bar{\mu}}
             dx^{\bar{\nu}}+\Omega_{\bar{i}\bar{j}}(x^M)
             \left(dx^{\bar{i}}-A_{\bar{\mu}}^{\bar{i}}(x^M)dx^{\bar{\mu}}
             \right)\left(dx^{\bar{j}}-A_{\bar{\nu}}^{\bar{j}}(x^M)
             dx^{\bar{\nu}}\right) \,,
\eequ
where we have split the space-time coordinates as
$x^M=(x^{\bar{\mu}},x^{\bar{i}})$; $M,N=0,\dots, D-1$ are space-time indices,
$\bar{\mu},\bar{\nu}=0,1, 2$ are three dimensional indices and
$\bar{i},\bar{j}=3,\dots D-1$ are indices in the remaining $D-3$ dimensions. We
may think of the space-time as a fibre bundle; $\{x^{\bar{i}}\}$ are coordinates
along the fibre and $\{x^{\bar{\mu}}\}$ are coordinates on the base space.

We are interested in studying $D$~dimensional space-times with an $SO(D-2)$
isometry group. This is the isometry group of the $S^{D-3}$ sphere, which
justifies why we are performing a $3+(D-3)$ splitting of the $D$~dimensional
space-time. Thus, we assume that $\xi_a$, $a=1,\dots, (D-3)(D-2)/2$, are Killing
vector fields,
\bequ
  \mathcal{L}_{\xi_a} \hat{g}_{MN}=0 \,,
  \label{kvf}
\eequ
with Lie algebra
\bequ
  \left[\xi_a,\xi_b\right]=\epsilon_{ab}{}^c\xi_c \, , 
  \label{algebra}
\eequ
where $\epsilon_{ab}{}^{c}$ are the structure constants of $SO(D-2)$. Because
the fibre has the minimal dimension necessary to accommodate $(D-3)(D-2)/2$
independent Killing vector fields, we may assume without loss of generality that
the Killing vector fields have components exclusively along the fibre:
$\xi_a=\xi_a^{\bar{i}}\partial_{\bar{i}}$.  Furthermore, we may normalise the
Killing vectors so that they only depend on the coordinates of the fibre,
\emph{i.e.\ }$\partial_{\bar\mu}\xi_a^{\bar i}=0$. Then Eq.~\eqref{kvf} gives
the following conditions
\begin{align}
  \label{Omega}
  \mathcal{L}_{\xi_a}  \Omega_{\bar{i}\bar{j}} & =0 \, , \\
  \label{comm}
  \mathcal{L}_{\xi_a}  A_{\bar{\mu}}^{\bar{i}} & =0 \, , \\
  \label{g}
  \mathcal{L}_{\xi_a}  g_{\bar{\mu}\bar{\nu}} & =0 \, .
\end{align}
These expressions can be interpreted either as Lie derivatives of
rank-$2$ tensors defined on the $D$~dimensional space-time, or as Lie
derivatives of a rank-$2$ tensor, a vector and a scalar, which are
defined on $S^{D-3}$.

Conditions \eqref{Omega}-\eqref{g} have the following implications:
\begin{enumerate}
\item[]
\bequ
  \Omega_{\bar{i}\bar{j}}=
      f(x^{\bar{\mu}})h_{\bar{i}\bar{j}}^{S^{D-3}} \, ,
\eequ
because, from \eqref{Omega},  $\Omega_{\bar{i}\bar{j}}$ admits the maximal number of Killing vector fields and thus must be the metric on a maximally
symmetric space at each $x^{\bar{\mu}}$.  Due to \eqref{algebra} this space must be the 
$S^{D-3}$ sphere.
$h_{\bar{i}\bar{j}}^{S^{D-3}}$ denotes the metric on an $S^{D-3}$ with unit
radius;
\item[]
\bequ
  g_{\bar{\mu}\bar{\nu}}=
      g_{\bar{\mu}\bar{\nu}}(x^{\bar{\mu}}) \, ,
\eequ
because the Killing vector fields $\xi_a$ act transitively on the fibre
and therefore the base space metric must be independent of the fibre
coordinates;

\item[]
\bequ
  A_{\bar{\mu}}^{\bar{i}}=0\, ,
  \label{novectors}
\eequ
because Eq.~(\ref{comm}) is equivalent to 
\begin{equation}
  [\xi_a,A_{\bar\mu}]=0\label{comm1} \ ,
\end{equation}
and there exist no non-trivial vector fields on $S^{D-3}$ for $D\ge 5$
that commute with all Killing vector fields on the sphere.

\end{enumerate}

We remark that \eqref{comm1} corresponds to the statement, expressed in
\cite{Cho:1986wk} in group theoretical language, that the gauge group for a
theory reduced on a coset space $G/H$ is the normaliser of $H$ in $G$; in the
case of a sphere, where $G=SO(D-2)$ and $H=SO(D-3)$, the normaliser vanishes and
then there are no ``gauge vectors'', \emph{i.e.}, no non-vanishing metric
components $g_{\bar\mu\bar i}$. If the normaliser of $H$ in $G$ is
non-vanishing, such metric components appear, and with dimensional reduction
they yield vector fields which contribute to the stress-energy tensor in the
reduced theory.  For example, in the case of head-on collision, if $D=4$, the
isometry space is $SO(2)$ and the quasi-matter of the reduced theory consists of
a scalar field and of a vector field (as in \cite{Geroch:1970nt} and in
\cite{Sjodin:2000zd,Sperhake:2000fe,Choptuik:2003as}); if $D>4$, the isometry
space is $SO(D-2)/SO(D-3)$, and the quasi-matter of the reduced theory consists
of a single scalar field.  In the remainder of this work we focus on this
subclass of space-times, which already contains a vast class of physically
relevant problems, and postpone a discussion of the general case
with $A_{\bar{\mu}}^{\bar{i}}\neq0$ (\emph{i.e.}, with $g_{\bar\mu\bar i}\neq0$)
to future work.

In practice, we are actually interested in performing a $4+(D-4)$ split of the
$D$ dimensional space-time. This may be done as follows. The metric on a unit
$S^{D-3}$ may always be written in terms of the line element on a unit
$S^{D-4}$, denoted by $d\Omega_{D-4}$, as follows,
\bequ
  h_{\bar{i}\bar{j}}^{S^{D-3}} dx^{\bar{i}}dx^{\bar{j}}=
      d\theta^2+\sin^2\theta d\Omega_{D-4} \, ,
\eequ
where $\theta$ is a polar-like coordinate, $\theta\in [0,\pi]$. Now we
introduce four dimensional coordinates, $x^\mu=(x^{\bar{\mu}},\theta)$,
$\mu=0,1,2,3$, and define a four dimensional metric
\bequ
  g_{\mu\nu}dx^\mu dx^\nu=g_{\bar{\mu}\bar{\nu}}dx^{\bar{\mu}}dx^{\bar{\nu}}
      +f(x^{\bar{\mu}}) d\theta^2 \, ,
   \label{4d}
\eequ
as well as a new conformal factor
\bequ
  \lambda(x^{\mu})=\sin^2\theta g_{\theta\theta} \, .
  \label{conformal}
\eequ
Then, the most general $D$~dimensional metric compatible with $SO(D-2)$
isometry is, for $D\ge 5$
\bequ
  d\hat{s}^2= g_{\mu\nu}dx^\mu dx^\nu + \lambda(x^{\mu}) d \Omega_{D-4} \, .
  \label{ansatz}
\eequ

Without specifying \eqref{4d} and \eqref{conformal}, the geometry \eqref{ansatz}
has only a manifest $SO(D-3)$ symmetry. We now perform a dimensional reduction
on a $(D-4)$-sphere.  This yields, from the $D$~dimensional vacuum Einstein
equations, a set of $3+1$ dimensional Einstein equations coupled to
quasi-matter. If $SO(D-2)$ is the full isometry group, the quasi-matter terms do
not contain independent degrees of freedom; rather, they may be completely
determined by the $3+1$ dimensional geometry, via \eqref{conformal}.  In this
case, we could perform a dimensional reduction on a $(D-3)$-sphere, which has
the full isometry group $SO(D-2)$.  This would yield a 2+1 dimensional
system. The former method allows, however, the use of existing numerical codes,
with small changes, which justifies our choice.

\begin{figure}[h!]


\centering
\includegraphics[width=0.8\textwidth]{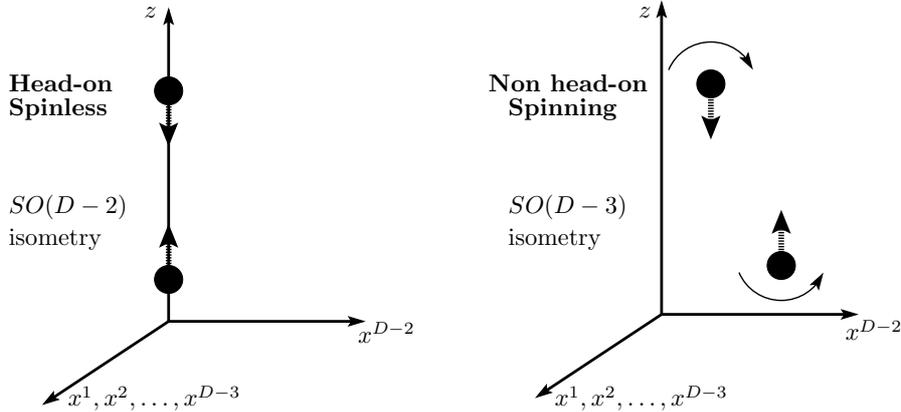}

\caption{$D$~dimensional representation, using coordinates $(t, x^1,x^2,\dots,
  x^{D-3},x^{D-2},z)$, of two types of BH collisions: (left panel) head-on for
  spinless BHs, for which the isometry group is $SO(D-2)$; (right panel) non
  head-on, with motion on a \textit{single} 2-plane, for BHs spinning in that
  \textit{same} plane only, for which the isometry group is $SO(D-3)$. The
  figures make manifest the isometry group in both cases.}
\label{headon_impact}
\end{figure}

The equations derived with dimensional reduction on a $(D-4)$-sphere can be
applied, of course, to describe also space-times in which the \emph{full}
isometry group is $SO(D-3)$.  This is the isometry group of a class of BH
collisions with impact parameter and with spin: the collisions in which the two
BHs always move on the same 2-plane and the only non trivial components of the
spin 2-form are on that same 2-plane -- see Fig.~\ref{headon_impact}.  With our
framework we are able, therefore, to describe not only head-on collisions of
spinless BHs but also a class of collisions for spinning BHs with impact
parameter.  As follows from the discussion of \eqref{novectors}, the ansatz
\eqref{ansatz} describes general space-times with $SO(D-3)$ isometry in $D\ge
6$.  We remark that the models with $D\ge6$ are actually the most interesting
for phenomenological studies of large extra dimensions models (see for instance
\cite{Kanti:2004nr}).

\subsection{Dimensional reduction on a $(D-4)$-sphere and $3+1$ split}
In the following we take \eqref{ansatz} as an ansatz, which has a manifest
$SO(D-3)$ isometry. The $D$~dimensional pure Einstein theory reduces then to a
four dimensional theory of gravity coupled to a scalar field
$\lambda(x^\mu)$. We remark that in this theory $\lambda$ and $g_{\mu\nu}$ are
viewed as independent degrees of freedom; the relations \eqref{4d},
\eqref{conformal} select a subset of the solution space. The solutions belonging
to this subset have enhanced isometry $SO(D-2)$ and correspond to some of the
physical processes we want to study (for instance, head-on collisions of
spinless BHs).

The $D$~dimensional Einstein-Hilbert action reduces to
\begin{equation}
  \mathcal{S} = \frac{1}{16\pi G_4}\int d^4x\sqrt{-g} \lambda^{\frac{D-4}{2}}
  \left[
    R + (D-4) \left(
      (D-5) \lambda^{-1} - \lambda^{-1} \Box \lambda
      - \frac{D-7}{4} \lambda^{-2} \partial_\mu \lambda \partial^\mu \lambda
    \right)
  \right] \, ,
\end{equation}
where the $D$~dimensional Newton's constant $G_D$ is related to the four
dimensional one $G_4$ by the area of the unit ${D-4}$ dimensional sphere:
$G_4=G_D/A^{S^{D-4}}$. Explicitly, the $D$~dimensional Einstein's equations in
vacuum yield the following system of four dimensional equations coupled to a
scalar field:
\begin{align}
  R_{\mu\nu} & =\frac{D-4}{2 \lambda}\left(\nabla_\mu\partial_\nu \lambda
      - \frac{1}{2\lambda} \partial_\mu\lambda \partial_\nu \lambda \right) \, ,
  \label{4deinstein} \\
  \nabla^\mu \partial_\mu \lambda & = 2(D-5)
      - \frac{D-6}{2\lambda} \partial_\mu \lambda \partial^\mu \lambda \, .
  \label{scalar}
\end{align}
In these equations, all operators are covariant with respect to the four
dimensional metric $g_{\mu\nu}$. The energy momentum tensor is\footnote{We use
  the standard form of the Einstein equations $G_{\mu\nu}=8\pi T_{\mu\nu}$ and
  choose geometrised units throughout.}
\bequ
  T_{\mu\nu}=\frac{D-4}{16\pi \lambda}\left[\nabla_\mu\partial_\nu \lambda
      - \frac{1}{2\lambda} \partial_\mu\lambda \partial_\nu\lambda
      - (D-5)g_{\mu\nu} + \frac{D-5}{4\lambda} g_{\mu \nu} \partial_\alpha
      \lambda \partial^\alpha \lambda \right] \, .
  \label{emtensor}
\eequ

With this four dimensional perspective, the usual $3+1$ split of space-time
\cite{Arnowitt:1962hi, York1979} can be performed (see, \textit{e.g.\
}\cite{Gourgoulhon:2007ue,Alcubierre:2008}).  For this purpose, we introduce the
projection operator $\gamma_{\mu\nu}$ and the normal to the three dimensional
hyper-surface $\Sigma$, $n^\mu$ ($n^\mu n_\mu = -1$),
\begin{align}
  \label{eq:2}
  \gamma_{\mu\nu} = g_{\mu\nu} + n_\mu n_\nu \, ,
\end{align}
as well as the lapse $\alpha$ and shift $\beta^\mu$,
\begin{align}
  \label{eq:4}
  \partial_t = \alpha n + \beta \ ,
\end{align}
where $t$ is the time coordinate. The four dimensional metric is then
written in the form
\bequ
  \label{4dinitial}
  ds^2=g_{\mu\nu}dx^\mu dx^\nu=-\alpha^2dt^2+\gamma_{ij}
      (dx^i+\beta^idt)(dx^j+\beta^jdt) \, , \qquad i,j=1,2,3 \, .
\eequ

As usual, we introduce the extrinsic curvature $K_{ij} = -\half
\mathcal{L}_n \gamma_{ij}$, which gives the evolution equation for
the $3$-metric,
\begin{align}
  \label{eq:gammaevol0}
  \left(
    \partial_t - \mathcal{L}_\beta
  \right)  \gamma_{ij} = - 2\alpha K_{ij} \, .
\end{align}

The time evolution for $K_{ij}$ is given by
\begin{align}
  \label{eq:Kevol00}
  \left(
    \partial_t -  \mathcal{L}_\beta
  \right) K_{ij} = -D_i \partial_j \alpha
  + \alpha \left(
    {}^{(3)}\! R_{ij} + K K_{ij} - 2K_{i k} K^{k}{}_j
  \right)
  - \alpha \gamma^\mu{}_i \gamma^\nu{}_j R_{\mu\nu} \, ,
\end{align}
where $D_i$ is the covariant derivative on the hyper-surface. The last
term, $ \gamma^\mu{}_i \gamma^\nu{}_j R_{\mu\nu}$, vanishes for vacuum
solutions. In the present case, it is given by the projection of equation
\eqref{4deinstein},
\begin{align}
  \gamma^\mu{}_i \gamma^\nu{}_j R_{\mu\nu} = \frac{D-4}{2 \lambda} \left(
      \gamma^\mu{}_i \gamma^\nu{}_j  \nabla_\mu\partial_\nu \lambda
    - \frac{1}{2\lambda} \partial_i\lambda \partial_j \lambda \right) \, .
\end{align}
Using the formula
\begin{align}
  \label{eq:proj}
  D_\alpha D_\beta \lambda
      = -K_{\alpha\beta} n^\sigma \partial_\sigma \lambda
      +  \gamma^\mu{}_\alpha \gamma^\nu{}_\beta \nabla_\nu \partial_\mu
      \lambda \, ,
\end{align}
and defining the variable
\begin{align}
  \label{eq:Kphi0}
  K_\lambda \equiv - \frac{1}{2} \mathcal{L}_n \lambda =
      - \frac{1}{2} n^\mu \partial_\mu \lambda \, ,
\end{align}
we obtain
\begin{align}
  \gamma^\mu{}_i \gamma^\nu{}_j \nabla_\nu \partial_\mu \lambda =
      D_i \partial_j \lambda - 2 K_{ij} K_\lambda \, .
\end{align}
Thus, \eqref{eq:Kevol00} becomes
\begin{equation}
  \label{eq:Kevol0}
  \begin{split}
    \left(
      \partial_t -  \mathcal{L}_\beta
    \right) K_{ij} & = -D_i \partial_j \alpha
    + \alpha \left(
      {}^{(3)}\! R_{ij} + K K_{ij} - 2K_{i k} K^{k}{}_j
    \right) \\
    &{}\quad - \alpha \frac{D-4}{2\lambda}  \left( D_i \partial_j \lambda
      - 2 K_{ij} K_\lambda - \frac{1}{2\lambda} \partial_i \lambda
      \partial_j \lambda \right)\, .
  \end{split}
\end{equation}
To summarise, the evolution equations for the 3-metric and extrinsic
curvature are \eqref{eq:gammaevol0} and \eqref{eq:Kevol0}.

If the isometry group is $SO(D-3)$, the quasi-matter field $\lambda$ represents
an independent degree of freedom, and we need to solve the evolution equations
for $\lambda$ and $K_\lambda$. Even in the case of the larger isometry
$SO(D-2)$, the evolution equations for $\lambda$ and $K_{\lambda}$ are useful as
they enable us to test Eq.~(\ref{conformal}) and thus provide a check of the
numerical evolution.  The evolution equation for $\lambda$ is \eqref{eq:Kphi0}
\bequ
  \label{phievo}
  \left(
    \partial_t - \mathcal{L}_\beta \right)
  \lambda  = - 2 \alpha K_\lambda \, .
\eequ
Eq. \eqref{scalar} provides an evolution equation for
$K_\lambda$. The contraction of Eq. \eqref{eq:proj} with $g^{\alpha\beta}$,
yields
\begin{align}
  \Box \lambda = \gamma^{ij}D_i\partial_j \lambda - 2 K K_\lambda
      - n^\mu n^\nu \nabla_\nu \partial_\mu \lambda \, .
\end{align}
Noting that
\begin{align}
  \mathcal{L}_n K_\lambda = n^\mu \partial_\mu K_\lambda 
  = -\half n^\mu \nabla_\mu n^\nu \partial_\nu \lambda
  - \half n^\mu n^\nu \nabla_\mu \partial_\nu \lambda \, ,
\end{align}
and 
\begin{align}
  n^\mu \nabla_\mu n^\nu = \frac{1}{\alpha} D^\nu \alpha \, ,
\end{align}
we obtain
\begin{align}
  - n^\mu n^\nu \nabla_\mu \partial_\nu \lambda =  2 \mathcal{L}_n K_\lambda
  + \frac{1}{\alpha} D^\nu \alpha \partial_\nu \lambda \, .
\end{align}
Noticing also that $D^\nu \alpha \partial_\nu \lambda = \gamma^{ij}\partial_i \alpha \partial_j
\lambda  $, we write
\begin{align}
  \Box \lambda = \gamma^{ij}D_i\partial_j \lambda - 2 K K_\lambda
      + 2 \mathcal{L}_n K_\lambda + \frac{1}{\alpha} \gamma^{ij}\partial_i \alpha \partial_j
\lambda   \, .
\end{align}
Moreover, from equation
\begin{align}
  D_\mu \lambda = \gamma^\nu{}_\mu \partial_\nu \lambda 
  = \partial_\mu \lambda - 2 n_\mu K_\lambda \, ,
\end{align}
we get
\begin{align}
  \partial_\alpha \lambda \partial^\alpha \lambda = \gamma^{ij}\partial_i \lambda \partial_j
       \lambda - 4 K_\lambda ^2 \, ,
\end{align}
so that the evolution equation for $K_\lambda$ is
\begin{align}
  \label{eq:evolphi0}
  \frac{1}{\alpha} \left(
    \partial_t - \mathcal{L}_\beta
  \right) K_\lambda
  = - \frac{1}{2\alpha} \gamma^{ij}\partial_i \lambda \partial_j \alpha
  + (D-5) + K K_\lambda + \frac{D-6}{\lambda} K_\lambda^2
  - \frac{D-6}{4\lambda} \gamma^{ij} \partial_i \lambda \partial_j \lambda
  - \half D^k \partial_k \lambda \, .
\end{align}
Equations \eqref{phievo} and \eqref{eq:evolphi0} are the evolution
equations for the quasi-matter degrees of freedom.

\subsection{BSSN formulation}\label{BSSN}
For numerical implementation, let us now write the evolution equations
in the Baumgarte, Shapiro, Shibata and Nakamura (BSSN) formulation
\cite{Shibata:1995we,Baumgarte:1998te}. Instead of evolving the variables
$\gamma_{ij}$ and $K_{ij}$, we introduce a conformal split of the physical
3-metric $\gamma_{ij}$ as
\bequ
  \gamma_{ij}\equiv \frac{1}{\chi}\tilde{\gamma}_{ij} \, .
\eequ
The conformal factor
\bequ
  \chi=\left({\rm det} \gamma_{ij}\right)^{-1/3} \, ,
\eequ
is chosen such that $\det\tilde{\gamma}_{ij}=1$ holds at all times. The
extrinsic curvature is split into a conformal traceless part,
$\tilde{A}_{ij}$, and its trace, $K$, as
\bequ
  \tilde{A}_{ij}\equiv \chi \left(K_{ij}-\frac{\gamma_{ij}}{3}K\right) \, .
\eequ
Moreover, we introduce the contracted conformal connection
\bequ
  \tilde{\Gamma}^i=\tilde{\gamma}^{jk}\tilde{\Gamma}^i_{jk} \, ,
\eequ
where
\begin{eqnarray}
   \Gamma^k_{ij} = \tilde{\Gamma}^k_{ij} - \frac{1}{2\chi}
      \left(\delta_i{}^k \partial_j \chi + \delta_j{}^k \partial_i \chi
      -\tilde{\gamma}_{ij}\tilde{\gamma}^{kl}\partial_l
      \chi \right) \ \Rightarrow \
 \Gamma^k = \chi \tilde{\Gamma}^k + \frac{1}{2}
      \tilde{\gamma}^{kl}\partial_l \chi \, ,
\end{eqnarray}
as an independent variable. In terms of the BSSN variables
$\chi,\tilde{\gamma}_{ij},\tilde{A}_{ij},\tilde{\Gamma}^k$, the evolution
equations are
\begin{subequations}
  \begin{align}
    \left( \partial_t -  \mathcal{L}_\beta \right) \tilde \gamma_{ij} & =
        - 2 \alpha \tilde A_{ij}\, , \\
    \left( \partial_t -  \mathcal{L}_\beta \right) \chi  & =
        \frac{2}{3} \alpha \chi K\, , \\
    \left( \partial_t -  \mathcal{L}_\beta \right) K & =
        [\dots] + 4 \pi \alpha (E + S)\, , \\
    \left( \partial_t -  \mathcal{L}_\beta \right) \tilde A_{ij} & =
        [\dots] - 8 \pi \alpha \left(
          \chi S_{ij} - \frac{S}{3} \tilde \gamma_{ij}
        \right)\, , \\
    \left( \partial_t -  \mathcal{L}_\beta \right) \tilde \Gamma^i & =
        [\dots] - 16 \pi \alpha \chi^{-1} j^i\, ,
  \end{align}
\end{subequations}
where $ [\dots] $ denotes the standard right-hand side of the BSSN equations in
the absence of source terms (see \emph{e.g.\ }\cite{Alcubierre:2008}); the
source terms are determined by
\begin{align}
  E & \equiv n^\alpha n^\beta T_{\alpha\beta} \, , \\
  j_i & \equiv - \gamma_i{}^\alpha n^\beta T_{\alpha \beta} \, , \\
  S_{ij} & \equiv \gamma^\alpha{}_i \gamma^\beta{}_j T_{\alpha \beta}\, , \\
  S & \equiv \gamma^{ij} S_{ij}\, ,
\end{align}
where the energy momentum tensor is given by Eq.~\eqref{emtensor}.
A straightforward computation shows that
\begin{subequations}
  \label{matterterms}
  \begin{align}
    \begin{split}
    \frac{4 \pi (E + S)}{D-4} & = -(D-5) \lambda^{-1} + \frac{1}{2}
        \lambda^{-1} \chi^{3/2} \tilde \gamma^{ij} \tilde D_i 
        \left(
          \chi^{-1/2} \partial_j \lambda
        \right)  \\
    &{} \quad
    + \frac{D-6}{4} \lambda^{-2} \chi \tilde \gamma^{ij} \partial_i
    \lambda \partial_j \lambda 
    - \lambda^{-1} K K_\lambda - (D-5) \lambda^{-2} K_\lambda^2 \, ,
  \end{split} \\
  \begin{split}
    \frac{8\pi \chi \left( S_{ij} - \frac{S}{3} \gamma_{ij}
      \right)}{D-4} & = \frac{1}{2} \chi \lambda^{-1} \tilde D_i
    \partial_j \lambda + \frac{1}{4} \lambda^{-1} \left(
      \partial_i \lambda \partial_j \chi + \partial_j
      \lambda \partial_i \chi - \tilde \gamma^{kl} \tilde
      \gamma_{ij} \partial_k \lambda
      \partial_l \chi \right) - \frac{1}{4} \chi
    \lambda^{-2} \partial_i
    \lambda \partial_j \lambda  \\
    &{} \quad - \lambda^{-1} K_\lambda \tilde{A}_{ij} -
    \frac{1}{6}\tilde \gamma_{ij} \lambda^{-1} \chi^{3/2} \tilde
    \gamma^{kl} \tilde D_k \left( \chi^{-1/2} \partial_l \lambda
    \right) + \frac{1}{12} \tilde \gamma_{ij} \lambda^{-2} \chi \tilde
    \gamma^{kl} \partial_l \lambda
    \partial_k \lambda \, ,
  \end{split} \\
%
        \frac{16 \pi \chi^{-1} j^i}{D-4} & = 2 \lambda^{-1}
        \tilde \gamma^{ij} \partial_j K_\lambda
        - \lambda^{-2} K_\lambda \tilde \gamma^{ij} \partial_j \lambda 
        - \tilde \gamma^{ik} \tilde \gamma^{lj} \tilde{A}_{kl} \lambda^{-1}
        \partial_j \lambda-\frac{\tilde{\gamma}^{ij}}{3}K\lambda^{-1}
        \partial_j\lambda  \, ,
  \end{align}
\end{subequations}
where $\tilde{D}_i$ is the covariant derivative with respect to
$\tilde{\gamma}_{ij}$.

Finally, the evolution equations for $\lambda$ and $K_\lambda$ are
\begin{subequations}
  \label{kl}
  \begin{align}
    \left( \partial_t - \mathcal{L}_\beta \right) \lambda & = - 2
        \alpha K_\lambda , \\
        \begin{split}
          \left( \partial_t - \mathcal{L}_\beta \right) K_{\lambda} &
          = \alpha \bigg\{ (D-5) + \frac{6-D}{4} \left[ \lambda^{-1}
            \chi \tilde{\gamma}^{ij}\partial_i \lambda
            \partial_j \lambda - 4 \lambda^{-1} K_\lambda^2
          \right] \\
          &{} \quad + K K_\lambda - \half \chi^{3/2} \tilde
          \gamma^{kl} \tilde D_k \left( \chi^{-1/2} \partial_l \lambda
          \right) \bigg\} - \half \chi \tilde{\gamma}^{ij}\partial_j
          \alpha \partial_i \lambda \, .
        \end{split}
  \end{align}
\end{subequations}
As stated before, in the case of head-on collisions of spinless BHs the full
symmetry of the $D$~dimensional system we want to consider makes
equations~\eqref{kl} redundant, by virtue of~\eqref{conformal}.  This allows to
determine the quasi-matter degree of freedom in terms of the three dimensional
spatial geometry, at each time slice.  Indeed, we have only used an $SO(D-3)$
subgroup in the dimensional reduction we have performed. The extra symmetry
manifests itself in the fact that $\gamma_{ij}$ possesses, at all times, (at
least) one Killing vector field. If one chooses coordinates adapted to this
Killing vector field, $\partial/\partial \theta$, the metric can then be written
in the form \eqref{4d}, and then the quasi-matter degree of freedom can be
determined from the spatial geometry by \eqref{conformal}. In the numerical
implementation, one can either determine, at each time-step, the scalar field
through (\ref{conformal}), or impose (\ref{conformal}) only in the initial data,
and then evolve the scalar field using Eq.~(\ref{kl}).

\section{Initial data}
\label{sec:initial-data}
Following the approach in \cite{Yoshino:2005ps,Yoshino:2006kc}, we now 
derive the initial data of the evolution.

\subsection{$D$~dimensional Hamiltonian and momentum constraints}
\label{sec:init-data-equations}
Let $\bar \Sigma$ be a $(D-1)$-dimensional space-like hyper-surface with
induced metric $\bar \gamma_{ab}$ and extrinsic curvature $\bar K_{ab}$
in the $D$~dimensional space-time.
The space-time metric has the form
\begin{align}
  d\hat{s}^2 = \hat g_{MN} dx^M dx^N = -\alpha^2 dt^2 
      + \bar \gamma_{ab} \left(
      dx^a + \beta^a dt
      \right) \left(
        dx^b + \beta^b dt
      \right),
  \label{metricinitial}
\end{align}
where lower case latin indices take values $a=1,\dots,D-1$.
The constraint equations are
\begin{align}
  \label{eq:hamiltonian2}
  &  \bar R + {\bar K}^2 - \bar K_{ab} {\bar K}^{ab} = 0 \, , \\
  \label{eq:momentum2}
  &  \bar D_a \left( \bar K^{ab} - \bar \gamma^{ab} \bar K  \right) = 0 \, ,
\end{align}
where $ \bar R$ is the Ricci scalar of the hyper-surface $\bar \Sigma$, $\bar K$
is the trace of the extrinsic curvature and $\bar D_a$ is the covariant
derivative with respect to $ \bar \gamma_{ab} $.

We conformally decompose the spatial metric
\begin{align}
  \label{eq:conformal}
  \bar  \gamma_{ab} & = \psi^{\frac{4}{D-3}} \hat \gamma_{ab}\, ,
\end{align}
which introduces the conformal factor $\psi$, and split the extrinsic curvature in trace and trace-free parts,
\begin{equation}
  \bar K_{ab} \equiv \bar A_{ab} + \frac{\bar K}{D-1} \bar \gamma_{ab}\, ,
\end{equation}
where $ \bar \gamma^{ab} \bar A_{ab} = 0$. Define $\bar A^{ab} \equiv
\bar \gamma^{ac} \bar \gamma^{bd} \bar A_{cd}$; define also the quantity
\begin{equation}
  \hat A^{ab}  \equiv  \psi^{2 \frac{D+1}{D-3}} \bar A^{ab} \, ,
\end{equation}
and lower its indices with $\hat \gamma_{ab}$,
\begin{equation}
  \hat A_{ab} \equiv \hat \gamma_{ac} \hat \gamma_{bd} \hat A^{cd} 
  = \psi^2 \bar A_{ab} \, .
\end{equation}
Assuming that the ``conformal metric'' $\hat \gamma_{ab}$ is flat, which is a
good approximation for the class of problems we want to study, we impose the
``maximal slicing condition'' $\bar K = 0 $. Then, the Hamiltonian and momentum
constraints become
\begin{align}
  & \hat\nabla_a \hat A^{ab} = 0\, , \label{eq:vacuum_hamilton} \\
  &  \hat \triangle \psi + \frac{D-3}{4(D-2)} \psi^{- \frac{3D -5}{D-3}   }
     \hat A^{ab} \hat A_{ab} = 0 \, ,
  \label{eq:vacuum_momentum}
\end{align}
where $\hat\nabla$ is the covariant derivative with respect to $\hat
\gamma_{ab}$ and $\hat \triangle$ is the flat space Laplace operator.

\subsection{Brill-Lindquist initial data and matching to four dimensions}
\label{sec:brill-lindq-init}
The simplest way to solve the constraints
\eqref{eq:vacuum_hamilton}-\eqref{eq:vacuum_momentum} is to require the
extrinsic curvature to be zero
\bequ
  \bar K_{ab}=0 \, .
  \label{initialD}
\eequ
This is sufficient to model the evolution of a single BH or even of $N$
non-spinning, non-boosted BHs. The constraints reduce to a simple harmonic
equation for the conformal factor, $\hat \triangle \psi =0$, which we solve in
cylindrical coordinates $\{x^a\}=(z,\rho,\theta, \dots)$, where `$\dots$'
represent the coordinates on the $(D-4)$-sphere,
\bequ
  \hat\gamma_{ab}dx^adx^b=dz^2+d\rho^2+\rho^2\left(d\theta^2+\sin^2\theta
       d\Omega_{D-4}\right) \, .
  \label{initialspatial}
\eequ
This choice of coordinates makes manifest the symmetries we want to
impose. Observe that $\theta$ is a polar rather than an azimuthal coordinate,
\emph{i.e.\ }$\theta\in [0,\pi]$. Next, we introduce ``incomplete'' Cartesian
coordinates as
\bequ
  x=\rho \cos\theta \, ,  \qquad y=\rho \sin\theta \, ,
  \label{inccartesian}
\eequ
where $-\infty<x<+\infty$ and $0\le y<+\infty$; we can then write the
$D$~dimensional initial data as \eqref{initialD} together with 
\bequ
  \bar \gamma_{ab}dx^adx^b=\psi^{\frac{4}{D-3}}\left[dx^2+dy^2+dz^2+y^2
      d\Omega_{D-4}\right] \, ,
  \label{smhigher}
\eequ
where $\psi$ is a harmonic function on  \eqref{initialspatial}.

If we compare the space-time metric \eqref{metricinitial} at the initial time
slice, for which the spatial metric is given by \eqref{eq:conformal}
and \eqref{smhigher}, with the generic form that has an $SO(D-3)$ symmetry and is
given by \eqref{ansatz}, \eqref{4dinitial}, we see that the initial data
for the four dimensional variables are
\bequ
  \label{k0}
  \gamma_{ij}dx^idx^j=\psi^{\frac{4}{D-3}}\left[dx^2+dy^2+dz^2\right] \, ,
\eequ
and
\bequ
  \lambda=y^2\psi^{\frac{4}{D-3}} \, .
  \label{iniscalar}
\eequ

It remains to determine the initial conditions for $K_{ij}$ and
$K_\lambda$. Using a set of $D$~dimensional coordinates that make manifest the
$SO(D-3)$ isometry, such as the one used in \eqref{smhigher}, the vanishing of
the extrinsic curvature $\bar{K}_{ij}$ is equivalent to
\bequ
  \label{k1}
  K_{ij}=0 \, ,
\eequ
whereas the vanishing of the components of $\bar K_{ab}$ along the
$(D-4)$-sphere implies that
\bequ
  \label{k2}
  K_\lambda=0 \, .
\eequ
Equations~\eqref{k0}--\eqref{k2} represent the Brill-Lindquist initial data
in our framework.

\subsubsection{Evolution of a single black hole}\label{single}
As one test of our framework we study the case of a single, non-spinning
BH. Even though the space-time is static, the slicing evolves when using the
puncture gauge.

The solution for the conformal factor, which shall be used in the numerical
tests to be presented below, is given by
\begin{align}
  \label{eq:psiBL}
  \psi \equiv 1 + \frac{\mu^{D-3}} {4 \left[x^2+y^2+(z-z_{BH})^2
      \right]^{(D-3)/2}} \, ,
\end{align}
where the ``puncture'' \cite{Brandt:1997tf} is placed at $x=y=0$ and
$z=z_{BH}$. In this formulation, there is an interesting signature that the BH
we wish to evolve is higher dimensional: the fall off of $\psi$, which is that
of a harmonic function in $D-1$ spatial dimensions.  Because the Tangherlini
solution \cite{Tangherlini:1963bw} may be expressed, in the same coordinate
system as used in \eqref{smhigher}, as
\bequ
  d\hat s^2=-\left(\frac{4R^{D-3}-\mu^{D-3}}{4R^{D-3}+\mu^{D-3}}\right)^2dt^2
      +\left(1+\frac{\mu^{D-3}}{4R^{D-3}}\right)^\frac{4}{D-3}
      \left(dx^2+dy^2+dz^2+y^2d\Omega_{D-4}\right)  \, ,
\eequ
where $R=\sqrt{x^2+y^2+z^2}$, we conclude that the parameter $\mu$ appearing in
the initial condition \eqref{eq:psiBL} is the same which appears in this form of
the Tangherlini solution. It is related to the ADM mass by
\bequ
  \mu^{D-3}=\frac{16\pi M_{ADM}}{(D-2)A^{S^{D-2}}} \, .
\eequ
Note, however, that this form of the Tangherlini solution is not appropriate for
a comparison with the numerical data. Indeed, the evolution does not, in
general, preserve the conformally flat slicing of the initial condition, which
is the slicing used in this form of the Tangherlini solution. We shall return to
this issue in Section~\ref{numres}.

\subsubsection{Head-on collision of black holes}
\label{sec:head-on-init}

As another test of our formulation, and in particular of the numerical code's
long term stability, we also evolve a head-on collision of non-spinning
non-boosted BHs.  In this case, the initial data for the conformal factor are
given by
\begin{align}
  \psi \equiv 1 + \frac{\mu_{\rm A}^{D-3}} {4 \left[x^2+y^2+(z-z_{\rm A})^2
      \right]^{(D-3)/2}}
    + \frac{\mu_{\rm B}^{D-3}} {4 \left[x^2+y^2+(z-z_{\rm B})^2
  \right]^{(D-3)/2}} \, .
\end{align}
This conformal factor is used in Section~\ref{numres}.

\section{The numerical treatment}
\label{sec:numerical-treatment}
Our numerical simulations have been performed by adapting the \textsc{Lean} code
\cite{Sperhake:2006cy}, initially designed for 3+1 vacuum space-times. The
\textsc{Lean} code is based on the \textsc{Cactus} computational toolkit
\cite{cactus}. It employs the BSSN formulation of the Einstein equations
\cite{Shibata:1995we,Baumgarte:1998te}, uses the moving puncture method
\cite{Campanelli:2005dd,Baker:2005vv}, the \textsc{Carpet} package for mesh
refinement \cite{Schnetter:2003rb,carpet}, the spectral solver described in
\cite{Ansorg:2004ds} for 3+1 initial data and Thornburg's {\sc AHFinderDirect}
\cite{Thornburg:1995cp,Thornburg:2003sf}.  Details about \textsc{Lean} may be
found in \cite{Sperhake:2006cy}. Here we focus on the numerical issues generated
by the quasi-matter terms arising from the dimensional reduction by isometry.

We expect that the quasi-matter field $\lambda$ has a $y^2$ fall off as
$y\rightarrow 0$, that is, on the $xz$ plane. This leads to divisions by zero on
the right-hand side of the BSSN evolution equations,
cf. \eqref{matterterms}. Since we expect all variables to remain regular on the
$xz$ plane, all divisions by $y$ need to be cancelled by a corresponding fall
off behaviour of the numerators. At $y=0$, however, in order to implement this
behaviour numerically, we need to isolate the irregular terms and evaluate
expressions such as
\begin{equation}
  \lim_{y\rightarrow 0} \frac{f}{y}\, ,
\end{equation}
where $f$ is some example function which behaves like $y^n$ with $n\ge 1$ near
the $xz$ plane. It is necessary, for this purpose, to formulate the equations in
terms of variables which are manifestly regular at $y=0$.  We also prefer to
apply a conformal re-scaling of $\lambda$ and use the evolution variable
\begin{equation}
\label{kappavariable}
  \zeta \equiv  \frac{\chi}{y^2} \lambda \, .
\end{equation}

As in \eqref{kl}, in order to obtain a first order evolution system in time, we
introduce an auxiliary variable (see Appendix~\ref{axis}):
\begin{eqnarray}
  K_{\zeta} \equiv  -\frac{1}{2\alpha y^2}(\partial_t-\mathcal{L}_\beta)
      (\zeta y^2)=-\frac{1}{2\alpha} \left( \partial_t \zeta
      - \beta^m \partial_m \zeta + \frac{2}{3}\zeta
      \partial_m \beta^m - 2\zeta \frac{\beta^y}{y} \right) \, .
  \label{eq:Kkappa_v2}
\end{eqnarray}
The third term on the right-hand side arises from the fact that $\zeta$ is not a
scalar, but a scalar density of weight $-2/3$.  The inclusion of this term might
not be necessary for a stable numerical implementation.  For consistency with
the rest of the BSSN variables, however, we decide to keep this form of
$K_{\zeta}$.

The quasi-matter terms \eqref{matterterms}, the quasi-matter evolution equations
\eqref{kl} and the constraints are recast in terms of $\zeta$ and $K_{\zeta}$ in
Appendix~\ref{axis}.  In particular, we notice that
\begin{eqnarray}
  K_{\lambda} = \frac{y^2}{\chi} K_{\zeta} + \frac{1}{3}
      \frac{y^2\zeta}{\chi} K\, .
      \label{useful}
\end{eqnarray}
A detailed analysis of the equations in terms of the variables $\zeta$ and
$K_{\zeta}$ shows how all terms with an explicit dependence on $1/y^n$, $n\ge 1$
may be treated for numerical implementation. This is discussed in
Appendix~\ref{troubleterms}.

\subsection{Numerical results in $D=5$}\label{numres}
We first address the question of longevity of our simulations in $D=5$. It is
also of interest in this context to test the code's capability to successfully
merge a BH binary. For this purpose we have evolved a head-on collision starting
from rest.  The initial conditions are those from Section~\ref{sec:head-on-init}
with
\begin{eqnarray}
  && \mu^2_{\rm A} = \mu^2_{\rm B} \equiv \frac{\mu^2}{2} \, , \\
  && z_A = -z_B = 3.185~\mu \, ,
\end{eqnarray}
and we use the grid setup (cf.~Sec.~II E of Ref.~\cite{Sperhake:2006cy})
\begin{equation}
  \left\{(512,~256,~128,~64,~32,~16,~8) \times (2,~1),~h=1/32\right\} \, , \nonumber
\end{equation}
in units of $\mu$. The gauge variables $\alpha$ and $\beta^i$ are evolved
according to the modified moving puncture conditions \eqref{eq:dtalpha} and
\eqref{eq:dtbeta} with parameters $\eta_K = \eta_{K_\zeta}=1.5$ and $\eta=0.75$.
We employ fourth order discretization in space and time and impose a floor value
\cite{Campanelli:2005dd} for the variable $\chi=10^{-4}$.

\begin{figure}[h!tb]
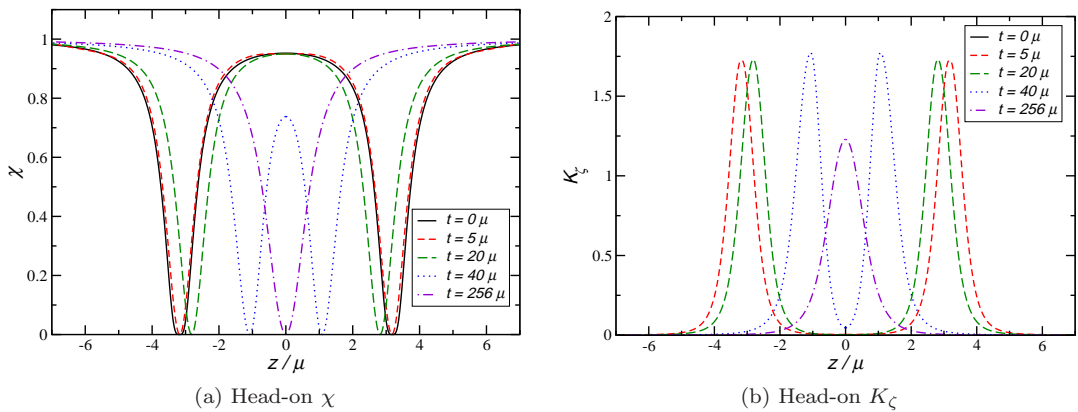

  \centering
  \subfloat[Head-on $\chi$]{
    \includegraphics[clip=true,width=0.46\textwidth]{headon_stable}
  }
\quad
  \subfloat[Head-on $K_{\zeta}$]{
    \includegraphics[clip=true,width=0.46\textwidth]{headon_stable_Kz}
  }
  \caption{The BSSN variable $\chi$ (left panel) and the quasi-matter momentum
           $K_{\zeta}$ (right panel) are shown along the axis of collision
           for a head-on collision at times $t=0$, $5$, $20$, $40$
           and $256~\mu$. Note that $K_{\zeta}=0$
           at $t=0$. }
  \label{headon_stable}
\end{figure}
In Fig.~\ref{headon_stable} we show the conformal factor $\chi$ and the momentum
$K_{\zeta}$ along the axis of collision at various times.  At early times, the
evolution is dominated by the adjustment of the gauge (cf.~the solid and
short-dashed curves). The two holes next start approaching each other
(long-dashed and dotted curves) and eventually merge and settle down into a
single stationary hole (dash-dotted curves). We have not observed any signs of
instability and decided to stop the simulation at $t=256~\mu$. It is reassuring
to notice that the framework can handle the merger in as robust a fashion as has
been demonstrated by various numerical groups for BH binaries in 3+1 dimensions.

We have also used the head-on collision to test the relation between the scalar
field $\lambda$ and the $3+1$ metric discussed in Sec.~\ref{sec:axial-symmetry}
for the case that $SO(D-2)$ is the full isometry group. We have verified for
this purpose that Eq.~\ref{conformal} remains satisfied to within a relative
error of $10^{-3}$ in the immediate vicinity of the puncture and at most
$10^{-5}$ everywhere else.

\begin{figure}[h!]
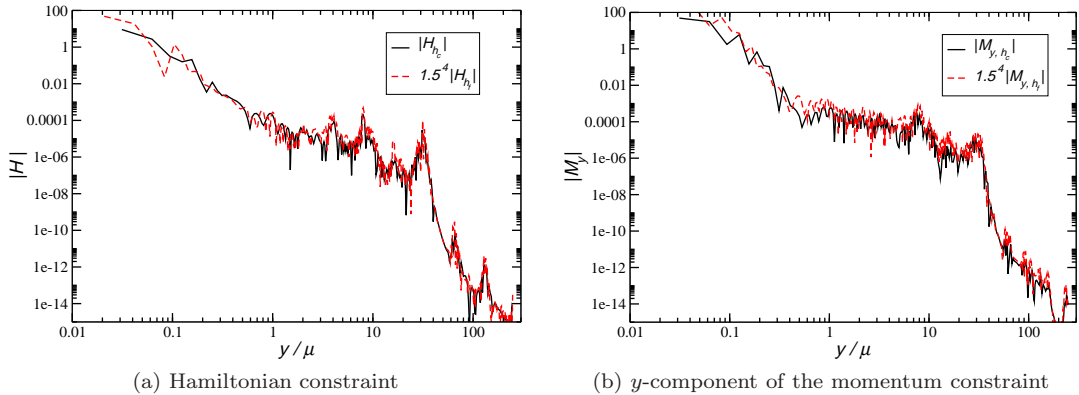

  \centering
  \subfloat[Hamiltonian constraint]{
    \includegraphics[clip=true,width=0.46\textwidth]{hamiltonian}
  }
\quad
  \subfloat[$y$-component of the
  momentum constraint]{
    \includegraphics[clip=true,width=0.46\textwidth]{momentum}
  }
  \caption{Constraints at time $t=28\mu$, for the evolution
  of a single Tangherlini BH in five dimensions.}
  \label{constraintsplot}
\end{figure}

In order to further test our numerical framework, we have performed simulations
of a single BH, using the initial data described in Section~\ref{single} and the
grid setup
\begin{equation}
  \left\{(512,~256,~128,~64,~32,~16,~8,~4,~2) \times (),~h\right\} \, ,\nonumber
\end{equation}
in units of $\mu$ with resolutions $h_{\rm c}=1/32$ and $h_{\rm f}=1/48$.  In
Fig.~\ref{constraintsplot} we show the Hamiltonian constraint and the $y$
component of the momentum constraint at evolution time $t=28\mu$. By this time
there is hardly any more gauge dynamics going on. One can see that there is some
noise, but the overall convergence is acceptable. For the Hamiltonian constraint
the convergence is essentially 4th order and for the momentum constraint it
decreases slightly towards 2nd or 3rd order in patches. From experience in 3+1
dimensional numerical relativity this is perfectly acceptable, especially given
the fact that prolongation in time is second-order accurate.

%
%
%

A different test of our numerical code was performed in order to compare the
analytical Tangherlini solution with our numerical results. The challenge to do
this comparison, at the level of the line element, is to write the well known
analytical solution in the same coordinate system in which the numerical
evolution is occurring.  One way around this problem is to fix the numerical
gauge as to match a known coordinate system for the analytic solution. Following
\cite{Yoshino:2009xp} we fixed the gauge parameters to be
\bequ
  \alpha=1 \, , \qquad \beta^i=0  \,  , \, i=1,2,3 \, ;
\eequ
this corresponds to \textit{geodesic slicing}. The $D$~dimensional Tangherlini
solution may be expressed in a coordinate system of type \eqref{metricinitial}
with $\alpha=1,\beta^a=0$, $a=1,\dots, D-1$. This coordinate system may be
achieved by setting a congruence of in-falling radial time-like geodesics, each
geodesic starting from rest at radial coordinate $r_0$, with $r_0$ spanning the
interval $[\mu,+\infty[$, and using their proper time $\tau$ and $r_0$ as
coordinates (instead of the standard $t$, $r$ Schwarzschild-like coordinates). A
detailed construction of the Tangherlini solution in these coordinates is given
in Appendix~\ref{geoslice}. The line element becomes
\bequ
  ds^2=-d\tau^2+\frac{\left(r_0(R)^2+\left(\frac{\mu}{r_0(R)}\right)^2
      \tau^2\right)^2}{r_0(R)^2-\left(\frac{\mu}{r_0(R)}\right)^2\tau^2}
      \frac{dR^2}{R^2}+\left(r_0(R)^2-\left(\frac{\mu}{r_0(R)}\right)^2
      \tau^2\right)d\Omega_3 \, ,
  \label{geodesicmetric}
\eequ
where $r_0(R)$ is given by Eq.~\eqref{rzerotor}.

\begin{figure}[h!]
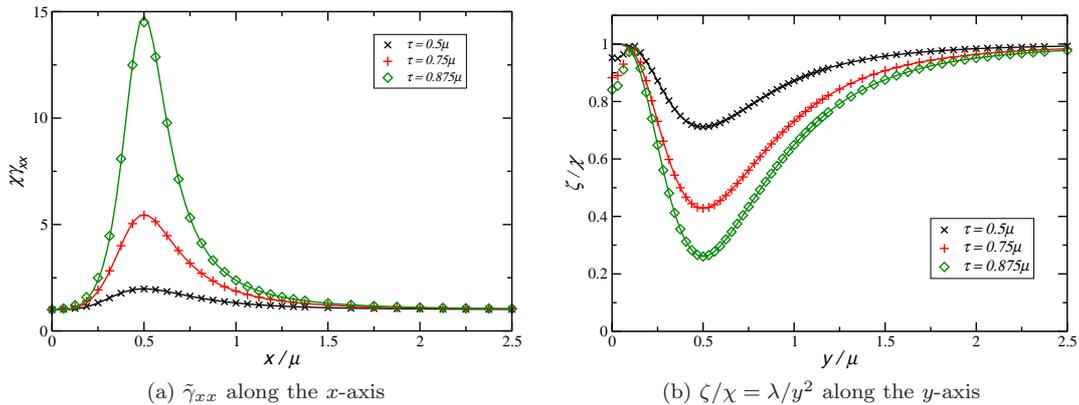

  \centering
  \subfloat[$\tilde{\gamma}_{xx}$ along the $x$-axis]{
    \includegraphics[clip=true,width=0.46\textwidth]{hxx}
  }
  \quad
  \subfloat[$\zeta/\chi=\lambda/y^2$ along the $y$-axis]{
    \includegraphics[clip=true,width=0.46\textwidth]{kappa}
  }
  \caption{Numerical values versus analytical plot (solid lines) 
           for various values of $\tau$, for
            the single Tangherlini BH
           in five dimensions. The horizontal axes are labelled in
           units of $\mu$.
  }
  \label{gammaxx}
\end{figure}

The numerical evolution in this gauge is naturally doomed. Geodesics hit the
physical singularity at finite proper time. Thus, this slicing is inappropriate
for a long term numerical evolution. As long as the evolution does not break
down, however, there is perfect control over the slicing, and hence the
numerical and analytical evolution can be compared with ease. This is shown in
Fig.~\ref{gammaxx}, where we have plotted one metric component
$\tilde{\gamma}_{xx}$ along the $x$ axis (left) and $\zeta/\chi$ (right), for
various values of $\tau$ using both the analytical solution and numerical
data. The agreement is excellent for $\tilde{\gamma}_{xx}$ and good for
$\zeta/\chi$. The latter shows some deviations very close to the puncture, but
we believe that it is not a problem for two reasons: \textit{(i)}~the agreement
improves for higher resolution; \textit{(ii)}~the mismatch does not propagate
outside of the horizon.

It is easy to interpret the behaviour observed for $\tilde{\gamma}_{xx}$. The
geodesic that starts from $r=r_0$ (in Schwarzschild-like coordinates) hits the
physical singularity of the Tangherlini solution within proper time
$\tau=r_0^2/\mu$. Moreover, this happens at
\bequ
  R=\frac{\mu}{2}\frac{1}{\sqrt{\tau/\mu}\pm\sqrt{\tau/\mu-1}} \, .
  \label{hitssing}
\eequ
The earliest time at which the slicing hits the singularity is $\tau=\mu$, which
happens at $R=\mu/2$. On the $x$-axis $R=x$ and indeed one sees in
Fig.~\ref{gammaxx} that $\tilde{\gamma}_{xx}$ diverges at $x=\mu/2$. The
divergence then extends to both larger and smaller values of $x$, as expected
from \eqref{hitssing}.

\subsection{Preliminary numerical results in $D=6$}\label{numres6}
A quick glance at the evolution equations (\ref{evotildek}) and
(\ref{evomomentumk}) of the scalar field $\zeta$ as well as the source terms
(\ref{eq:ESnew})-(\ref{eq:jinew}) indicates that $D=5$ may be a special case. In
all these expressions there exist terms which manifestly vanish for $D=5$.  In
contrast, there exist no terms which manifestly vanish for any dimension $D \ge
6$.  The purpose of this Section is to extend the test of our framework to a
case which involves all source terms.

We have indeed noticed one fundamental difference between simulations in $D=5$
and those using $D\ge6$. Whereas we have been able to obtain stable simulations
of single BHs lasting hundreds of $\mu$ for the former case by modifying the
moving puncture gauge conditions, we have not yet succeeded in doing so for
$D\ge 6$. While the lifetime of the simulations in $D\ge 6$ shows a dependence
on the exact nature of lapse and shift, all simulations developed instabilities
on a timescale of about $10~\mu$. Resolving this issue requires an extensive
study involving a large number of experiments with gauge conditions, constraint
damping and possibly other aspects of the formulation. Such a study is beyond
the scope of this work and deferred to a future publication. The results
presented in this Section still provide valuable information. Most importantly,
they demonstrate the internal consistency of the code for $D\ge 6$ and thus
minimise the possibility of a simple error in the implementation.  Furthermore
they exhibit clearly that our framework and in particular our regularisation of
the variables as discussed in Appendix~\ref{troubleterms} is in principle
suitable for simulations in arbitrary dimensions.

We first consider the convergence of the constraints analogous to the results
displayed in Fig.~\ref{constraintsplot} for $D=5$. Compared to those
simulations, the only change we have applied in $D=6$ is to set the gauge
parameters to $\eta_K=\eta_{K_\zeta}= \eta=2$. This choice enables us to evolve
single BHs to about $10~\mu$ when instabilities cause the runs to abort.
\begin{figure}[htb]
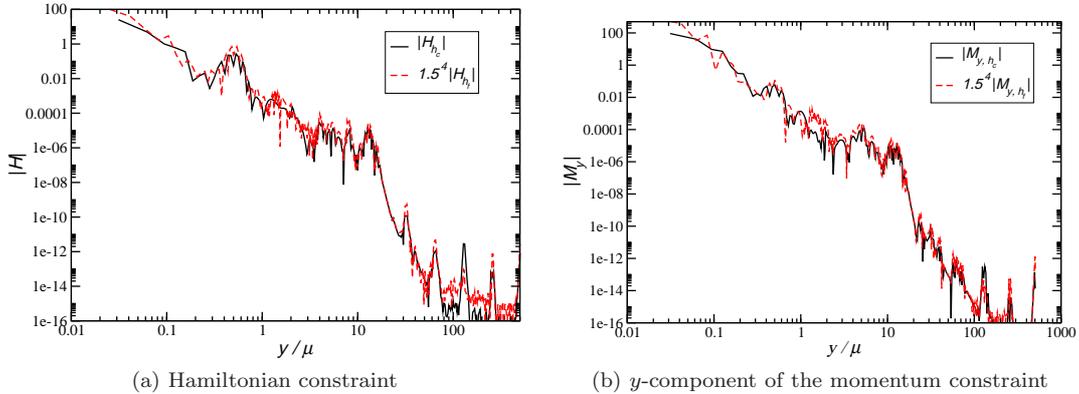

  \centering
  \subfloat[Hamiltonian constraint]{
    \includegraphics[clip=true,width=0.46\textwidth]{hamd6_t08}
  }
  \quad
  \subfloat[$y$-component of the momentum constraint]{
    \includegraphics[clip=true,width=0.46\textwidth]{momd6_t08}
  }
  \caption{Constraints at time $t=8\mu$, for the evolution
  of a single Tangherlini BH in six dimensions.}
  \label{constraintsd6}
\end{figure}
In Fig.~\ref{constraintsd6} we show the Hamiltonian and the $y$-component of the
momentum constraint at $t=8~\mu$ along the $y$-axis. As for $D=5$, the high
resolution result is amplified by a factor $1.5^4$ expected for fourth order
convergence \cite{Alcubierre:2008}. While the convergence appears to be closer
to second order in some patches of the momentum constraint, the results are
clearly compatible with the numerical discretization.

For the second test, we compare the numerical evolution of a single $D=6$
Tangherlini BH with the analytic solution, using geodesic slicing. This
comparison is more difficult in the present case than in $D=5$,
\begin{figure}[htb]
  \centering\includegraphics[clip=true,width=0.46\textwidth]{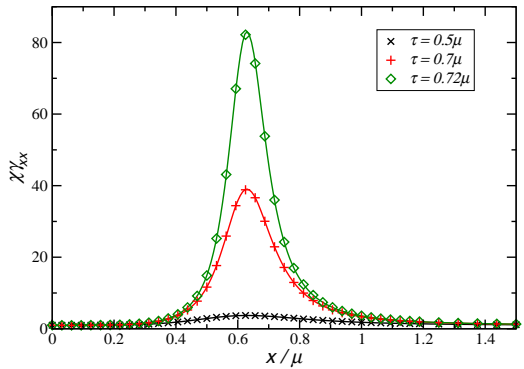}
  \caption{Numerical values versus the semi-analytic
           expression of $\tilde{\gamma}_{xx}$ (cf.~Appendix~\ref{geoslice})
           along the $x$-axis for the single Tangherlini BH in six dimensions.
  }
  \label{geodesic_d6_hxx}
\end{figure}
because the line element analogous to \eqref{geodesicmetric} cannot be obtained
in a simple analytic form. In Appendix~\ref{geoslice} we demonstrate how a
semi-analytic solution can be obtained for the metric. In
Fig.~\ref{geodesic_d6_hxx} we compare this expression with the three dimensional
numerical values at times $\tau=0.5~\mu$, $0.7~\mu$ and $0.72~\mu$. The
agreement is excellent and demonstrates that our code works well at least up to
the point where instabilities set in. As mentioned above, resolving these
stability problems will be of the highest priority in future extensions of our
work.

\section{Final Remarks}
\label{sec:final-remarks}

In this paper we present a framework that allows the generalisation of the
present generation of 3+1 numerical codes to evolve, with relatively minor
modifications, space-times with $SO(D-2)$ symmetry in $5$ dimensions and
$SO(D-3)$ symmetry in $D\ge 6$ dimensions.  The key idea is a dimensional
reduction of the problem along the lines of Geroch's \cite{Geroch:1970nt}
procedure that recasts the $D$~dimensional Einstein vacuum equations in the form
of the standard four dimensional equations plus some quasi-matter source terms.
The resulting equations can be transformed straightforwardly into the BSSN
formulation that has proved remarkably successful in numerical evolutions of BH
configurations in 3+1 space-times.  We have isolated several issues related to
the regularisation of the variables used in our formulation and demonstrated how
all difficulties related to the coordinate singularity arising out of the use of
a ``radius-like'' coordinate can be successfully addressed in a numerical
implementation. We have further illustrated how initial data for single,
non-spinning BHs as well as BH binaries with vanishing initial extrinsic
curvature can be adapted straightforwardly to the formulation presented in this
paper.  More generally, the class of problems that may be studied with our
framework includes head-on collisions in $D\ge 5$ and a subset of BH collisions
with impact parameter and spin in $D\ge 6$.

As might be expected, stable evolutions of such space-times require some
modifications of the underlying methods of the so-called {\em moving puncture}
technique, especially with regard to the gauge conditions used therein. We have
successfully modified the slicing condition via incorporation of the canonical
momentum of the quasi-matter field in order to obtain long-term stable
simulations in $D=5$ dimensions. Unfortunately, these modifications do not
appear sufficient to provide long-term stability for arbitrary values of the
dimensionality $D$.  We will address this important issue in the form of a
systematic study in future work.

We have tested our framework by adapting the {\sc Lean} code and performed a
variety of single BH space-times. Most importantly, we have demonstrated the
internal consistency of our numerical framework in $D=5$ and $6$ dimensions by
showing convergence of the Hamiltonian and momentum constraints as well as
comparing numerical results with (semi-)analytic expressions for a single
Tangherlini BH in geodesic slicing. We have further shown for $D=5$ that the
head-on collision of a BH binary successfully merges into a single hole which
settles down into a stationary state and can be evolved numerically for long
times, hundreds of $\mu$ in the present example.

A complete study of such BH binary evolutions requires the implementation of
gravitational wave extraction in arbitrary dimensions as well as the
generalisation of apparent horizon diagnostics beyond $D=4$. Both are currently
being implemented in the {\sc Lean} code and will be discussed in detail in
future work.

In spite of several open questions, we believe that our formalism will open up a
vast range of uncharted territory in BH physics for contemporary numerical
relativity.  The list of possible applications and extensions of our framework
is too large to be included here, and we merely mention strong hyperbolicity
studies of the BSSN formulation with sources and systematic investigation of BH
binary dynamics in $D$~dimensions.  These studies are under way and will be
reported elsewhere.

\begin{acknowledgments}
  We would like to thank L. Lindblom and M. Sampaio for discussions.  We also
  thank the participants of the V Iberian Cosmology Meeting, the XII Marcel
  Grossmann Meetings, the Spanish Relativity Meeting and the I and II Black
  Holes Workshop for useful feedback.  M.Z. and H.W. are funded by FCT through
  grants SFRH/BD/43558/2008 and SFRH/BD/46061/2008.  V.C. acknowledges financial
  support from Funda\c c\~ao Calouste Gulbenkian through a short-term
  scholarship.  V.C. and C.H. are supported by a ``Ci\^encia 2007'' research
  contract.  A.N. is funded by FCT through grant SFRH/BPD/47955/2008.  This work
  was partially supported by FCT - Portugal through projects
  PTDC/FIS/64175/2006, PTDC/FIS/098025/2008, PTDC/FIS/098032/2008
  PTDC/CTE-AST/098034/2008, CERN/FP/109306/2009, CERN/FP/109290/2009 as well as
  NSF grants PHY-090003, PHY-0900735, PHY-0601459, PHY-0652995
  and the Fairchild foundation to Caltech.
  Computations were performed on the TeraGrid clusters ranger and kraken and at
  Magerit in Madrid.
  The authors thankfully acknowledge the computer resources, technical
  expertise and assistance provided by the Barcelona Supercomputing Centre ---
  Centro Nacional de Supercomputaci\'on.
\end{acknowledgments}

\appendix
\section{Implementing regular variables}
\label{axis}
The numerical evolution faces a problem at the symmetry axis, given the
quasi-matter terms in \eqref{matterterms} and the initial data discussed in
Sec.~\ref{sec:initial-data}. The ``incomplete'' Cartesian coordinate $y$
vanishes at the symmetry axis, cf. \eqref{inccartesian}. Then, from
\eqref{iniscalar}, $\lambda$ vanishes at the axis (except, possibly, at the
puncture). Inspection of equations \eqref{matterterms} and \eqref{kl}
immediately reveals various divisions by $\lambda$, leading to numerical
problems.

From previous experience with polar and spherical coordinates in simpler models
involving, for example, neutron stars (cf.~\cite{Seidel:1990xb, Gabler:2009yt})
we know that it is better to avoid the use of singular variables such as
$\lambda$. We should use, instead, regular functions.  In our case, since
$\lambda$ behaves as $y^2$ near the axis, this is simply achieved by introducing
a variable $\zeta$ via \eqref{kappavariable}. The evolution of $\zeta$ is
formulated in terms of a first order in time system of equations. For this
purpose we have introduced in Eq.~(\ref{eq:Kkappa_v2}) the variable $K_{\zeta}$.
We remark that if, instead, we employ the standard definition for the momentum
associated with $\zeta$, \emph{i.e.}
\begin{eqnarray}
  \hat{K}_{\zeta} \equiv  -\frac{1}{2\alpha}(\partial_t-\mathcal{L}_\beta) \zeta=-\frac{1}{2\alpha} \left( \partial_t \zeta
      - \beta^m \partial_m \zeta + \frac{2}{3}\zeta
      \partial_m \beta^m  \right) \, ,
  \label{eq:Kkappa_v3}
\end{eqnarray}
we face problems in the numerical evolution for vanishing lapse. This may be
seen as follows. From \eqref{useful}
\begin{eqnarray}
  K_{\lambda} = \frac{y^2}{\chi} \hat{K}_{\zeta} + \frac{1}{3}
      \frac{y^2\zeta}{\chi} K+\frac{\beta^y}{\alpha}\frac{y\zeta}{\chi}\, .
\end{eqnarray}
Acting on both sides of this equation with the derivative operator
\begin{equation}
   \partial_0 \equiv \partial_t - \beta^m \partial_m \nonumber \, , 
\end{equation}
results in the expression
\begin{equation}
  \begin{split}
  \partial_0  \hat{K}_{\zeta} & = 
      \frac{\chi}{y^2} \partial_0
      K_{\lambda} + 4\frac{\beta^y}{y}  \hat{K}_{\zeta}
      + \frac{4}{3}\alpha K \hat{K}_{\zeta}
      + \frac{4}{3}\zeta \frac{\beta^y}{y}K
      + \frac{2}{9}\zeta \alpha K^2 \\
   & \quad - \frac{\zeta}{3} \partial_0 K
      + \frac{\zeta}{{\alpha}} \left( \frac{\beta^y}{y} \right)^2
      - \frac{2}{3}  \hat{K}_{\zeta} \partial_m \beta^m
      - \frac{\zeta}{\alpha}
      \frac{\partial_0 \beta^y}{y}
      + \zeta\frac{\beta^y}{y \alpha^2} \partial_0 \alpha
       \, .
       \label{trouble2}
     \end{split}
\end{equation}
This is an evolution equation for $\hat{K}_\zeta$. To obtain it explicitly one uses  \eqref{eq:evolphi0} to express $\partial_0 K_\lambda$, together with 
\begin{align}
  \partial_0 K & = -D^m \partial_m \alpha + \alpha
      \left( \tilde{A}^{mn} \tilde{A}_{mn} +\frac{1}{3} K^2\right)
      + 4\pi \alpha (E+S)\, .
\end{align}
Moreover we need gauge conditions. Throughout this work we use
the following coordinate choices
\begin{align}
  \partial_0 \alpha &= -2\alpha (\eta_K K + \eta_{K_\zeta} K_{\zeta})\, ,
      \label{eq:dtalpha} \\
  \partial_0 \beta^i &= \frac{3}{4} \tilde{\Gamma}^i - \eta \beta^i\, .
      \label{eq:dtbeta}
\end{align}
Note the extra term involving $K_{\zeta}$ in the slicing condition compared with
standard moving puncture gauge in 3+1 dimensions and the additional freedom we
have introduced in the form of the parameters $\eta_K$ and $\eta_{K_\zeta}$.

The problems in the case of a collapsed lapse become clear if we consider the
final two terms in \eqref{trouble2}.  These terms do not change when BSSN
variables are introduced and diverge for the modified moving puncture gauge
conditions (\ref{eq:dtalpha}) and (\ref{eq:dtbeta}) as the lapse $\alpha
\rightarrow 0$.  We have solved this problem by expressing our equations in
terms of the variable $K_\zeta$ \eqref{eq:Kkappa_v2}, instead of
$\hat{K}_{\zeta}$.

In BSSN variables, the evolution equation for $\zeta$ and $K_{\zeta}$ (which
replace the quasi-matter evolution equations \eqref{kl}) become
\begin{subequations}
  \label{finalmatter0}
  \begin{align}
  \partial_t \zeta &= -2\alpha K_{\zeta}
      + \beta^m \partial_m \zeta
      - \frac{2}{3} \zeta \partial_m \beta^m
      + 2\zeta \frac{\beta^y}{y}\, , 
      \label{evotildek} \\
      \begin{split}
        \partial_t K_{\zeta} & = \beta^m \partial_m K_{\zeta} -
        \frac{2}{3}K_{\zeta} \partial_m \beta^m +2\frac{\beta^y}{y}
        K_{\zeta} -\frac{1}{3} \zeta\partial_0 K - \frac{\chi
          \zeta}{y} \tilde{\gamma}^{ym} \partial_m \alpha -
        \frac{1}{2} \tilde{\gamma}^{mn} (\partial_m \alpha)
        (\chi \partial_n \zeta - \zeta \partial_n \chi)
        \\
        & \quad + \alpha \left[ \vphantom{\frac{\tilde{\Gamma}^y}{y}}
          (5-D)\frac{\chi}{y^2}(\zeta\tilde{\gamma}^{yy}-1)
          +(4-D)\frac{\chi}{y} \tilde{\gamma}^{ym} \partial_m \zeta
          +\frac{2D-7}{2} \frac{\zeta}{y} \tilde{\gamma}^{ym}
          \partial_m \chi
        \right. \\
        & \quad + \frac{6-D}{4} \frac{\chi}{\zeta} \tilde{\gamma}^{mn}
        (\partial_m \zeta)(\partial_n \zeta) + \frac{2D-7}{4}
        \tilde{\gamma}^{mn} (\partial_m \zeta) (\partial_n \chi) +
        \frac{1-D}{4} \frac{\zeta}{\chi} \tilde{\gamma}^{mn}
        (\partial_m \chi) (\partial_n \chi)
        \\
        & \quad + (D-6) \frac{K_{\zeta}^2}{\zeta} + \frac{2D - 5}{3}
        KK_{\zeta} + \frac{D-1}{9} \zeta K^2 + \frac{1}{2}
        \tilde{\gamma}^{mn} \left( \zeta \tilde{D}_m \partial_n \chi -
          \chi \tilde{D}_m \partial_n \zeta \right) \left. +\chi \zeta
          \frac{\tilde{\Gamma}^y}{y} \right] \, .
        \label{evomomentumk}
      \end{split}
  \end{align}
\end{subequations}
These equations have no manifest problems as $\alpha\rightarrow 0$.

In terms of the regular variables, $\zeta$ and $K_{\zeta}$,
the quasi-matter terms \eqref{matterterms} read
\begin{subequations}
  \label{finalmatter}
  \begin{align}
    \begin{split}
      \frac{4\pi(E+S)}{D-4} & = (D-5) \frac{\chi}{\zeta}
      \frac{\tilde{\gamma}^{yy} \zeta - 1}{y^2} -\frac{2D-7}{4\tilde
        \kappa}\tilde{\gamma}^{mn} (\partial_m\zeta) (\partial_n \chi)
      - \chi \frac{\tilde{\Gamma}^y}{y} 
      \\
      & \quad +\frac{D-6}{4} \frac{\chi}{\zeta^2} \tilde{\gamma}^{mn}
      (\partial_m \zeta)(\partial_n \zeta) +\frac{1}{2\zeta}
      \tilde{\gamma}^{mn} (\chi \tilde{D}_m \partial_n \zeta - \zeta
      \tilde{D}_m \partial_n \chi) 
        \\
      & \quad -\frac{K K_{\zeta}}{\zeta}
      - \frac{1}{3}K^2+(D-4) \frac{\tilde{\gamma}^{ym}}{y}
      \left( \frac{\chi}{\zeta}
        \partial_m \zeta - \partial_m \chi \right) - \frac{1}{2}
      \frac{\tilde{\gamma}^{ym}}{y} \partial_m \chi 
         \\
         & \quad + \frac{D-1}{4} \tilde{\gamma}^{mn}
         \frac{(\partial_m \chi) (\partial_n \chi)}
         {\chi}   
         -(D-5) \left(\frac{K_{\zeta}}{\zeta} +
           \frac{K}{3}\right)^2 \, ,
         \label{eq:ESnew}
    \end{split} \\
    \begin{split}
      \frac{8\pi \chi \left( S_{ij} - \frac{1}{3} \gamma_{ij}
          S\right)}{D-4} & = \frac{1}{2} \left[ \frac{\chi}{y\zeta} \left(
          \delta_j{}^y \partial_i \zeta + \delta_i{}^y \partial_j \zeta -
          2\zeta \tilde{\Gamma}^y_{ij} \right) + \frac{1}{2\chi}
        (\partial_i \chi) (\partial_j \chi) + \frac{\chi}{\zeta}
        \tilde{D}_i \partial_j \zeta
      \right.    \\
      & \quad \left. - \tilde{D}_i \partial_j \chi +
        \frac{1}{2\chi}\tilde{\gamma}_{ij} \tilde{\gamma}^{mn}
        \partial_n \chi \left(\partial_m \chi -
          \frac{\chi}{\zeta} \partial_m \zeta \right) -
        \tilde{\gamma}_{ij} \frac{\tilde{\gamma^{ym}}}{y} \partial_m \chi
      \right.   \\
      & \quad \left. - \frac{\chi}{2\zeta^2} (\partial_i \zeta)
        (\partial_j \zeta) \vphantom{\frac{\chi}{y\zeta}} \right]^{\rm TF}
      - \left( \frac{K_{\zeta}}{\zeta} + \frac{1}{3}K \right)
      \tilde{A}_{ij} \, ,
      \label{eq:Sij_v2}
    \end{split} \\
    \begin{split}
      \frac{16\pi j_i}{D-4} & = \frac{2}{y} \left[ \delta_i{}^y
        \frac{K_{\zeta}}{\zeta} - \tilde{\gamma}^{ym} \tilde{A}_{mi}
      \right] +2\frac{1}{\zeta}\partial_i K_{\zeta} -
      \frac{K_{\zeta}}{\zeta} \left( \frac{1}{\chi}\partial_i\chi +
        \frac{1}{\zeta}
        \partial_i \zeta \right)  \\
      & \quad + \frac{2}{3} \partial_i K - \tilde{\gamma}^{nm}
      \tilde{A}_{mi} \left( \frac{1}{\zeta} \partial_n \zeta
        -\frac{1}{\chi} \partial_n \chi \right)\, .
      \label{eq:jinew}
    \end{split}
  \end{align}
\end{subequations}

Finally, the constraints are now given by
\begin{align}
  \mathcal{H} & \equiv 
  R + \frac{2}{3}K^2 - \tilde{\gamma}^{mn}\tilde{\gamma}^{kl}
  \tilde{A}_{mk} \tilde{A}_{nl} - 16\pi E\, , \label{hamiltonianfinal}\\
  \mathcal{M}_i & \equiv 
  \tilde{\gamma}^{mn} \left( \tilde{D}_n \tilde{A}_{im}
      - \frac{3}{2}\tilde{A}_{mi} \frac{\partial_n \chi}{\chi} \right)
      - \frac{2}{3} \partial_i K -8\pi j_i \, , \label{momentumfinal}
\end{align}
where we also need to express $E$ in terms of our fundamental
variables. It is given by
\begin{equation}
\begin{split}
  \frac{16\pi E}{D-4} & = (D-3) \frac{\chi}{y\zeta} \tilde{\gamma}^{ym}
       \partial_m \zeta
       - (D-2) \frac{1}{y}\tilde{\gamma}^{ym} \partial_m \chi
       + \frac{D-7}{4}
       \frac{\chi}{\zeta^2} \tilde{\gamma}^{mn}(\partial_m
       \zeta)(\partial_n \zeta) \\
    & \quad - \frac{D-2}{2\zeta} \tilde{\gamma}^{mn} (\partial_m
       \zeta)(\partial_n \chi)
       + \frac{D+3}{4\chi} \tilde{\gamma}^{mn} (\partial_m \chi)
       (\partial_n \chi)
       - (D-5) \frac{K_{\zeta}^2}{\zeta^2}
       -\frac{2D-4}{3} K \frac{K_{\zeta}}{\zeta} \\
    & \quad - \frac{D+1}{9}K^2
       + \frac{\chi}{\zeta} \tilde{\gamma}^{mn} \tilde{D}_m \partial_n
       \zeta - \tilde{\gamma}^{mn} \tilde{D}_m \partial_n \chi
       - 2 \chi \frac{\tilde{\Gamma}^y}{y} +(D-5) \frac{\chi}{\zeta}
       \frac{\tilde{\gamma}^{yy} \zeta - 1}{y^2}
       \, .
\end{split}
\end{equation}

\section{Analysis of troublesome terms at $y=0$} \label{troubleterms}
The right-hand sides of Eqs.~(\ref{finalmatter0})-(\ref{momentumfinal}) contain
various terms which cannot be evaluated directly at $y=0$ because they involve
explicit division by $y$.  Although these terms are regular by virtue of a
corresponding behaviour of the numerators, they need to be explicitly evaluated
in the numerical implementation. In this Appendix we outline how the regularity
of these terms can be implemented in a simple and efficient manner. For
convenience we use a special notation: late latin indices $i,~j,~\ldots$ run
from 1 to 3, covering $x$, $y$ and $z$, but early latin indices $a,~b,~\ldots$
take values 1 and 3 but not 2, \emph{i.e.\ }they cover $x$ and $z$ but not $y$.

We begin this discussion by describing a simple manipulation which underlies
most of our regularisation procedure. Consider for this purpose a function $h$
which is linear in $y$ near $y=0$, \emph{i.e.\ }its Taylor expansion is given by
$h(y) = h_1 y + \mathcal{O}(y^2)$. From this relation we directly obtain
\begin{equation}
  \lim_{y \rightarrow 0} \frac{h}{y} = h_1 = \partial_y h \, .
      \label{eq: yforderiv}
\end{equation}
This trading of divisions by $y$ for partial derivatives extends to higher
orders in a straightforward manner and will be used throughout the following
discussion.

Next, we consider the right-hand sides of
Eqs.~(\ref{finalmatter0})-(\ref{momentumfinal}) and summarise the potentially
troublesome terms as follows
\begin{align}
  &\frac{\beta^y}{y} \, , \qquad \frac{\tilde{\Gamma}^y}{y} \, ,
      \label{eq: irrbyoy} \\
  &\frac{\tilde{\gamma}^{ym}}{y}\partial_m f \, ,
      \label{eq: irrdf} \\
  &\frac{\tilde{\gamma}^{yy} \zeta - 1}{y^2} \, ,
      \label{eq: irrconsing} \\
  &\frac{1}{y} \left( \delta_i{}^y \frac{K_{\zeta}}{\zeta}
      - \tilde{\gamma}^{ym} \tilde{A}_{mi} \right) \, ,
      \label{eq: irrdtconsing} \\
  &\frac{1}{y} \left( \delta_j{}^y \partial_i \zeta + \delta_i{}^y \partial_j
      \zeta - 2\zeta \tilde{\Gamma}^y_{ij} \right) \, .
      \label{eq: irrChris}
\end{align}
Here $f$ stands for either of the scalars or densities $\zeta$, $\chi$ and
$\alpha$.

Regularity of the terms (\ref{eq: irrbyoy}) immediately follows from the
symmetry condition of the $y$-component of a vector
\begin{equation}
  \beta^y(-y) = -\beta^y(y) \, .
\end{equation}
We can therefore use the idea illustrated in Eq.~(\ref{eq: yforderiv})
and obtain
\begin{equation}
  \lim_{y\rightarrow 0} \frac{\beta^y}{y} = \partial_y \beta^y \, ,
\end{equation}
and likewise for $\tilde{\Gamma}^y/y$.
The terms (\ref{eq: irrdf}) are treated in a similar manner because
the derivative of a scalar (density) behaves like a vector on our
Cartesian grid. We thus obtain
\begin{equation}
  \lim_{y \rightarrow 0} \left( \frac{\tilde{\gamma}^{ym}}{y}
        \partial_m f \right)
      = (\partial_y \tilde{\gamma}^{ya})(\partial_a f)
        + \tilde{\gamma}^{yy} \partial_y \partial_y f \, .
\end{equation}

Regularity of the expression (\ref{eq: irrconsing}) is not immediately obvious
but can be shown to follow directly from the requirement that there should be no
conical singularity at $y=0$. Specifically, this condition implies that
$\tilde{\gamma}^{yy} \zeta = 1 + \mathcal{O}(y^2)$, so that
\begin{equation}
  \lim_{y \rightarrow 0} \left( \frac{\tilde{\gamma}^{yy} \zeta-1}{y^2} \right)
      = \frac{1}{2} \left( \zeta \partial_y \partial_y \tilde{\gamma}^{yy}
        + \tilde{\gamma}^{yy} \partial_y \partial_y \zeta \right) \, .
  \label{eq: regES_v3}
\end{equation}

The discussion of the term (\ref{eq: irrdtconsing}) requires us to distinguish
between the cases $i=a\ne y$ and $i=y$. The former straightforwardly results in
\begin{equation}
  \lim_{y\rightarrow 0} \left( -\frac{\tilde{\gamma}^{ym}}{y} \tilde{A}_{ma}
       \right)
     = -\tilde{A}_{ba} \partial_y \tilde{\gamma}^{yb}
       - \tilde{\gamma}^{yy} \partial_y \tilde{A}_{ya} \, .
\end{equation}
For the case $i=y$, we first note that the limit $y \rightarrow0$ implies
$\tilde{\gamma}^{yy} = 1/\tilde{\gamma}_{yy} + \mathcal{O}(y^2)$, so that the
condition (\ref{eq: regES_v3}), \emph{i.e.\ }no conical singularities, can be
written as
\begin{equation}
  \lim_{y \rightarrow 0} \left(\zeta - \tilde{\gamma}_{yy}\right)
      = \mathcal{O}(y^2) \, .
\end{equation}
Next we take the time derivative of this expression and obtain after some
manipulation
\begin{equation}
  \mathcal{O}(y^2) =
  \lim_{y \rightarrow 0} \partial_t(\zeta - \tilde{\gamma}_{yy})
     = -2\alpha \zeta \left( \frac{K_{\zeta}}{\zeta}
       - \tilde{\gamma}^{ym} \tilde{A}_{my} \right)
       + \mathcal{O}(y^2) \, ,
\end{equation}
and, consequently,
\begin{equation}
  \lim_{y \rightarrow 0} \left[ \frac{1}{y} \left( \frac{K_{\zeta}}{\zeta}
        - \tilde{\gamma}^{ym} \tilde{A}_{my} \right) \right]
      = 0 \, .
\end{equation}

Finally, we consider the term (\ref{eq: irrChris}). Expansion of the Christoffel
symbol, repeated use of the method illustrated in Eq.~(\ref{eq: yforderiv}) and
the condition for avoiding a conical singularity enable us to regularise this
term for all combinations of the free indices $i$ and $j$. We thus obtain
\begin{align}
  \lim_{y \rightarrow 0} \left[
      \frac{1}{y} \left( 2\partial_y \zeta - 2\zeta \tilde{\Gamma}^y_{yy}
      \right) \right]
    & = 2\partial_y \partial_y \zeta - \zeta \tilde{\gamma}^{yy}
      \partial_y \partial_y \tilde{\gamma}_{yy}
      - \zeta (\partial_y \tilde{\gamma}^{yc}) (2\partial_y \tilde{\gamma}_{yc}
      - \partial_c \tilde{\gamma}_{yy}) \, , \\
  \lim_{y \rightarrow 0} \left[
      \frac{1}{y} \left( \partial_a \zeta - 2\zeta \tilde{\Gamma}^y_{ay}
      \right) \right]
    &=  0 \, , \\
    \begin{split}
      \lim_{y \rightarrow 0} \left[
      -2\frac{\zeta}{y} \tilde{\Gamma}^y_{ab} \right]
    & = -\zeta \tilde{\gamma}^{yy} (\partial_y \partial_a \tilde{\gamma}_{by}
      +\partial_y \partial_b \tilde{\gamma}_{ya} - \partial_y \partial_y
      \tilde{\gamma}_{ab}) 
      \\    
      & \quad 
      - \zeta (\partial_y \tilde{\gamma}^{yc}) (\partial_a \tilde{\gamma}_{bc}
      + \partial_b \tilde{\gamma}_{ac} - \partial_c \tilde{\gamma}_{ab}) \, .
    \end{split}
\end{align}
We conclude this discussion with a method to express derivatives of the inverse
metric in terms of derivatives of the metric. For this purpose we use the
condition that $\det \tilde{\gamma}_{ij}=1$ by construction and explicitly
invert the metric components as for example in
\begin{equation}
  \tilde{\gamma}^{xy} = \tilde{\gamma}_{xz}\tilde{\gamma}_{yz}
       - \tilde{\gamma}_{xy} \tilde{\gamma}_{zz} \, .
\end{equation}
A straightforward calculation gives us the derivatives of the inverse metric
components as follows
\begin{align}
  \partial_y \tilde{\gamma}^{xy}
    & = \tilde{\gamma}_{xz} \partial_y \tilde{\gamma}_{yz}
       - \tilde{\gamma}_{zz} \partial_y \tilde{\gamma}_{xy}
       + \mathcal{O}(y^2) \, , \\
  \partial_y \tilde{\gamma}^{yz} & = \tilde{\gamma}_{xz} \partial_y
      \tilde{\gamma}_{xy} - \tilde{\gamma}_{xx} \partial_y
      \tilde{\gamma}_{yz} + \mathcal{O}(y^2)\, , \\
  \partial_y \tilde{\gamma}^{yy} & = \tilde{\gamma}_{zz} \partial_y
      \tilde{\gamma}_{xx} + \tilde{\gamma}_{xx} \partial_y
      \tilde{\gamma}_{zz} - 2\tilde{\gamma}_{xz} \partial_y
      \tilde{\gamma}_{xz} \, , \\
  \partial_y \partial_y \tilde{\gamma}^{yy} &=
      \tilde{\gamma}_{zz} \partial_y \partial_y \tilde{\gamma}_{xx}
      + \tilde{\gamma}_{xx} \partial_y \partial_y \tilde{\gamma}_{zz}
      - 2\tilde{\gamma}_{xz} \partial_y \partial_y \tilde{\gamma}_{xz}
      + \mathcal{O}(y^2) \, .
      \label{eq:dydygammayy}
\end{align}
The benefit in using these expressions is purely numerical: we do not need to
store the inverse metric in grid functions which reduces the memory requirements
of the simulations.

\section{Geodesic slicing}
\label{geoslice}
In standard Schwarzschild-like coordinates, the Tangherlini metric reads
\bequ
  ds^2=-f(r)dt^2+\frac{dr^2}{f(r)}+r^2d\Omega_{D-2} \, ,
      \qquad f(r)=1-\left(\frac{\mu}{r}\right)^{D-3}  \, .
\eequ
For a radially in-falling massive particle, starting from rest at $r=r_0$, the
energy per unit mass is $\sqrt{f(r_0)}$. The geodesic equation may then be
written as
\bequ
  \frac{dt}{d\tau}=\frac{\sqrt{f(r_0)}}{f(r)} \, ,
      \qquad \left(\frac{dr}{d\tau}\right)^2=f(r_0)-f(r) \, .
      \label{geodesics}
\eequ
In four and five dimensions these equations have simple solutions. In
five dimensions the solutions are
\bequ
  t=\sqrt{f(r_0)}\tau +\frac{\mu}{2}\ln \left|\frac{\tau+\sqrt{f(r_0)}
       r_0^2/\mu}{\tau-\sqrt{f(r_0)} r_0^2/\mu}\right| \, , \qquad r^2=r_0^2
       -\left(\frac{\mu}{r_0}\right)^2\tau^2 \, .
\eequ
Then, performing a coordinate transformation $(t,r)\rightarrow (\tau,
r_0)$ the line element becomes
\bequ
  ds^2=-d\tau^2+\frac{\left(r_0^2+\left(\frac{\mu}{r_0}\right)^2
      \tau^2\right)^2}{r_0^2-\left(\frac{\mu}{r_0}\right)^2
      \tau^2}\frac{dr_0^2}{r_0^2f(r_0)}+\left(r_0^2-\left(\frac{\mu}{r_0}
      \right)^2\tau^2\right)d\Omega_3 \, .
\eequ
This coordinate system encodes a space-time slicing with zero shift and constant
(unit) lapse (\emph{i.e.\ }of type \eqref{metricinitial} with
$\alpha=1,\beta^a=0$) for \textit{all} times. To compare it with a numerical
evolution we must have the initial data for the spatial metric written in a
conformally flat form. Taking the initial hyper-surface to be $\tau=0$ we see
that this is achieved by a coordinate transformation $r_0\rightarrow R$ with
\bequ
  \frac{dR}{R}=\frac{dr_0}{\sqrt{f(r_0)} r_0} \ \ \Rightarrow \ \ r_0(R)=
      R\left(1+\frac{\mu^2}{4R^2}\right) \, .
  \label{rzerotor}
\eequ
This actually coincides with the standard coordinate transformation from
Schwarzschild to isotropic coordinates in five dimensions. The line element
finally reads \eqref{geodesicmetric}. At the initial hyper-surface $\tau=0$,
\bequ
  ds^2_{\tau=0}=\left(\frac{r_0(R)}{R}\right)^2\left(dR^2+R^2d\Omega_3\right)=
      \left(\frac{r_0(\sqrt{\rho^2+z^2})}{\sqrt{\rho^2+z^2}}\right)^2
      \left(dz^2+d\rho^2+\rho^2d\theta^2+\rho^2\sin^2\theta d\Omega_1 \right)\, ,
\eequ
where we have used the metric on the 3-sphere in the form
\bequ
  d\Omega_3=d\tilde{\theta}+\sin^2\tilde{\theta}(d\theta^2+\sin^2\theta
       d\Omega_1) \, , \
  \label{S3}
\eequ
and performed the coordinate transformation $(R,\tilde{\theta})\rightarrow
(\rho,z)$ defined as
\bequ
  \rho=R\sin\tilde{\theta} \, , \ \ z=R\cos\tilde{\theta}  \, .
\eequ
Using \eqref{inccartesian} we get
\bequ
  ds^2_{\tau=0}= \left(\frac{r_0(\sqrt{x^2+y^2+z^2})}{\sqrt{x^2+y^2+z^2}}
      \right)^2\left(dx^2+dy^2+dz^2+y^2d\Omega_1\right)\, .
\eequ
Thus the coordinate transformation from the spherical coordinates used
in \eqref{geodesicmetric}, $(R,\tilde{\theta},\theta)$, to the ``incomplete''
Cartesian coordinates used in the numerical evolution $(x,y,z)$ is
\bequ
  x=R\sin\tilde{\theta}\cos\theta \, , \ \ y=R\sin\tilde{\theta}
      \sin\theta \, , \ \ z=R\cos\tilde{\theta} \, ,
  \label{3dinccart}
\eequ
which resembles the usual coordinate transformation from spherical polar
coordinates to Cartesian coordinates in $\mathbb{R}^3$; but note that
$\tilde{\theta}$ and $\theta$ are \textit{both} polar angles with range
$[0,\pi]$, which is the manifestation of the Cartesian coordinates
``incompleteness''.

The coordinate change \eqref{3dinccart} brings the five dimensional
Tan\-gher\-li\-ni metric in geode\-sic slicing to a conformally flat form at
$\tau=0$. This matches the initial data for the numerical evolution. One may
ask, however, if the coordinate transformation \textit{evolves}, in order to
compare the analytic form with the numerical evolution. This cannot be the case,
since the existence of $\tau$-dependent terms in the coordinate transformation
would imply a drift away from geodesic slicing. We are thus guaranteed that the
coordinate transformation \eqref{3dinccart} is valid for \textit{all} values of
$\tau$. Then, we can predict the value of the metric components that should be
obtained from the numerical evolution; say $\gamma_{xx}$ should be, at time
$\tau$
\bequ
  \gamma_{xx}(\tau,x,y,z)=\frac{x^2g_{RR}(\tau,R)}{R^2}+
      \frac{x^2z^2g_{\tilde{\theta}\tilde{\theta}}(\tau,R)}{R^4(x^2+y^2)}
      +\frac{y^2g_{\theta\theta}(\tau,R)}{(x^2+y^2)^2} \, ,
      \label{generalgammaxx}
\eequ
where $R^2=x^2+y^2+z^2$ and
$g_{RR}(\tau,R),g_{\tilde{\theta}\tilde{\theta}}(\tau,R),
g_{\theta\theta}(\tau,R)$ are readily obtained from \eqref{geodesicmetric} with
\eqref{S3} and \eqref{3dinccart}. The result for $\tilde{\gamma}_{xx}$ along the
$x$-axis is plotted in Fig.~\ref{gammaxx} for various values of $\tau$.

For $D\ge 6$ the situation is more involved because equations \eqref{geodesics}
can no longer be integrated straightforwardly, but require a numerical
treatment.  First one notices that the coordinate transformation
$(t,r)\rightarrow (\tau,r_0)$, with initial conditions $t(\tau=0)=0$ and
$r(\tau=0)=r_0$, brings the $D$~dimensional Tangherlini metric to the form
\bequ
  ds^2=-d\tau^2+\left(\frac{\partial r(\tau,r_0)}{\partial r_0}\right)^2
    \frac{dr_0^2}{f(r_0)}+r^2(\tau,r_0)d\Omega_{D-2} \, .
\eequ
Then, from the initial conditions, it follows that the coordinate
transformation to isotropic coordinates at $\tau=0$ is
\bequ
  \frac{dR}{R}=\frac{dr_0}{\sqrt{f(r_0)} r_0} \ \ \stackrel{D=6}{\Rightarrow} \ \ r_0(R)=\frac{
      R}{\mu}\left(1+\frac{\mu^3}{4R^3}\right)^{2/3} \, .
  \label{rzerotor6}
\eequ
Writing the metric on the $(D-2)$-sphere as in \eqref{S3} (replacing
$d\Omega_1\rightarrow d\Omega_{D-4}$), one concludes that the transformation to
``incomplete'' Cartesian coordinates is still \eqref{3dinccart}. Thus
\eqref{generalgammaxx} is still valid, which reduces to, along the $x$-axis
($R=x$):
\bequ
\gamma_{xx}(\tau,x,0,0)=g_{RR}(\tau,x)=\frac{r_0(x)^2}{x^2}
    \left(\frac{\partial r(\tau,r_0)}{\partial r_0}\right)^2_{r_0=r_0(x)} \, .
    \label{finalgs}
\eequ
This expression is valid for any $D$. For $D=6$, $r_0(x)$ is explicitly given by
\eqref{rzerotor6}. The derivative in \eqref{finalgs} has to be computed
numerically. The result for $\tilde{\gamma}_{xx}$ is plotted, for various values
of $\tau$, in Fig.~\ref{geodesic_d6_hxx}.


\bibliographystyle{myutphys}
\bibliography{num-rel}

\end{document}